\documentclass[a4paper,man,natbib]{apa6}\usepackage[]{graphicx}\usepackage[table]{xcolor}
\makeatletter
\def\maxwidth{ %
  \ifdim\Gin@nat@width>\linewidth
    \linewidth
  \else
    \Gin@nat@width
  \fi
}
\makeatother

\definecolor{fgcolor}{rgb}{0.345, 0.345, 0.345}

\usepackage{framed}
\makeatletter
 {\par\unskip\endMakeFramed%
 \at@end@of@kframe}
\makeatother

\definecolor{shadecolor}{rgb}{.97, .97, .97}
\definecolor{messagecolor}{rgb}{0, 0, 0}
\definecolor{warningcolor}{rgb}{1, 0, 1}
\definecolor{errorcolor}{rgb}{1, 0, 0}
\newenvironment{knitrout}{}{} % an empty environment to be redefined in TeX

\usepackage{alltt}

\usepackage[english]{babel}
\usepackage[utf8x]{inputenc}
\usepackage{amsmath}
\usepackage{graphicx}
\usepackage[colorinlistoftodos]{todonotes}
\usepackage{hyperref} % for clickable links
\usepackage{lineno}   % for line numbering
\usepackage{color}    % for changing colors
\usepackage{multirow} % for tables with more rows
\usepackage[table]{xcolor}

\newcommand{\pathBIB}{./bib}

\title{Solving the ``many variables'' problem in MICE with principal component regression}
\shorttitle{[PREPRINT] Solving the ``many variables'' problem}
\fourauthors
	{Edoardo Costantini}
	{Kyle M. Lang}
	{Klaas Sijtsma}
	{Tim Reeskens}
\fouraffiliations
	{Tilburg University, Department of Methodology and Statistics}
	{Utrecht University, Department of Methodology and Statistics}
	{Tilburg University, Department of Methodology and Statistics}
	{Tilburg University, Department of Sociology}
\authornote{
  Corresponding author's email address: \\
  e.costantini@tilburguniversity.edu; \\
  This article has been submitted to a journal but it was published on https://arxiv.org before formal peer review.
  }

\abstract{
Multiple Imputation (MI) is one of the most popular approaches to addressing missing values in questionnaires and surveys.
MI with multivariate imputation by chained equations (MICE) allows flexible imputation of many types of data.
In MICE, for each variable under imputation, the imputer needs to specify which variables should act as predictors in the imputation model.
The selection of these predictors is a difficult, but fundamental, step in the MI procedure, especially when there are many variables in a data set.
In this project, we explore the use of principal component regression (PCR) as a univariate imputation method in the MICE algorithm to automatically address the ``many variables'' problem that arises when imputing large social science data.
We compare different implementations of PCR-based MICE with a correlation-thresholding strategy through two Monte Carlo simulation studies and a case study.
We find the use of PCR on a variable-by-variable basis to perform best and that it can perform closely to expertly designed imputation procedures.}

\IfFileExists{upquote.sty}{\usepackage{upquote}}{}
\begin{document}
	\maketitle

    % Start section header numbering here
    \setcounter{secnumdepth}{3} % up to depth 3

    % Paper content
    
% Project:  paper-mipcr-compare
% Topic:    Introduction
% Author:   Edoardo Costantini
% Created:  2021-12-17
% Modified: 2023-01-09

\section{Introduction}\label{sec:introduction}

Missing values are a problem afflicting virtually all data sets in the social and behavioral sciences.
Multiple Imputation (MI) is one of the most popular approaches to address the issue of non-response.
Although MI can treat essentially any missing data problem, it was originally designed to impute large surveys, especially when those surveys are used to create publicly released data that many researchers will analyze independently \citep{rubin:1996}.
In this context, MI was envisioned as being especially useful when the data collector (and imputer) is distinguished from the ultimate user (or analyst).

The imputer's main task is to define an imputation model that supports analyses from many users.
A well-designed imputation model should include all the predictors of missingness present in the data, and it should incorporate all the features of the substantive analysis models that will be used by the data analysts.
If important predictors of missingness are left out of the imputation model, the missing at random (MAR) assumption is violated~\citep[p. 339]{collinsEtAl:2001}.
If some features of the substantive analysis model of interest do not appear in the imputation model, the two models are said to be uncongenial, a situation that can invalidate the inferential conclusions obtained after imputation (\citealp{meng:1994};~\citealp[p. 218]{littleRubin:2002}.)

To decide which predictors to include in the imputation model, a commonly recommended strategy is to include as many predictors as possible \citep[i.e., the inclusive strategy, ][]{collinsEtAl:2001}.
However, the scale of modern social surveys and data collection endeavors complicates the task of selecting these predictors.
Cross-sectional social surveys (e.g., World values survey, \citealp{WVS:2020}; European values study, \citealp{EVS:2017}) commonly measure hundreds of variables, and including all of these variables in the imputation model can lead to prohibitively long imputation times and convergence failures \citep[p. 259]{vanBuuren:2018}.
Social and behavioral scientists also frequently work with longitudinal surveys and panel studies (e.g., Panel Study of Income Dynamics, \citealp{mcgonagleEtAl:2012}; LISS Panel, \citealp{LISS:2018}), which can lead to data sets with many more columns than rows.
This high-dimensionality can result in singularity issues \citep[p. 46]{hastieEtAl:2009} when estimating the imputation models.
The imputer needs to address this ``many variables'' problem by thoroughly scanning all the available variables to decide which of them should be used in the imputation models.
In this article, we explore the use of principal component regression to automate the definition of the imputation model by replacing a large number of possible predictors with a small subset of principal components (PCs).

\subsection{MICE and the ``many variables'' problem}

In social science research, multivariate imputation by chained equation~\citep[MICE][]{mice} has been implemented in all major statistical software (e.g., Stata, \citealp{statacorp:2013}; SPSS, \citealp{IBM:2020}; R, \citealp{mice}) and is arguably the most popular way to implement MI\@.
MICE, also known as fully conditional specification and sequential regression imputation \citep{raghunathanEtAl:2001}, is an iterative algorithm that obtains imputations from the implied multivariate distribution of the missing data by sampling from a set of univariate conditional densities.
This algorithm requires the definition of a conditionally specified univariate imputation model for each variable under imputation.
At every iteration, each univariate imputation model is used to obtain replacement values for the missing data points.
When convergence is reached, any sample from the univariate imputation models' predictions represents a sample from the target multivariate data distribution.
These samples are used to define multiple versions of the original data, with different plausible values used to replace the original missing data.
Any analysis model of scientific interest can then be estimated on each of the multiply imputed datasets.
The estimates of the parameters of interest in the analysis model are then pooled following Rubin's rules~\citep[p. 76]{rubin:1987}.

The definition of the univariate imputation models is a fundamental step for the good performance of the MICE procedure.
For each variable under imputation, the imputer needs to define a univariate imputation model.
This task involves two decisions:
\begin{enumerate}
    \item Selecting the model form;
    \item Selecting the predictors.
\end{enumerate}

The first decision is usually guided simply by the measurement level of the variables under imputation.
For example, continuous variables can be imputed using a linear regression model, while binary variables can be imputed using logistic regression.
The second decision requires choosing the variables to be included as predictors in the imputation model.
In general, it is advisable to adopt an inclusive strategy~\citep{collinsEtAl:2001}, meaning including as many predictors as possible in the univariate imputation models.
Using as much information as possible from the data leads to multiple imputations that have minimal bias and maximal efficiency~\citep{meng:1994, collinsEtAl:2001}.
Furthermore, including more predictors in the univariate imputation models makes the MAR assumption more plausible~\citep[p. 339]{collinsEtAl:2001}.
Finally, if the imputation model omits variables that are part of the analysis model that will be estimated on the imputed data, the analysis model's parameter estimates might be biased \citep[p. 229]{enders:2010}, and the attendant confidence intervals might be too wide \citep[p. 218]{littleRubin:2002}.
As a result, including more predictors in the imputation models increases the range of analysis models that can be estimated with a given set of imputations~\citep{meng:1994}.

When a data set consists of only a few variables (i.e., tens of variables), it may be feasible to include all of these variables in all the univariate imputation models.
However, standard imputation methods face computational limitations in the presence of a large number of predictors (i.e., hundreds).
For example, MICE using Bayesian imputation under the normal linear model~\citep[p. 68]{vanBuuren:2018} requires the number of predictors ($p$) in the univariate imputation model to be smaller than the number of observed cases ($n$) to avoid computational problems with the system of equations~\citep[p. 203]{jamesEtAl:2013}.
Even when the number of predictors is smaller than the number of observations, including many predictors in the imputation models can increase the chances of collinearity issues~\citep[pp. 167--170]{vanBuuren:2018} and can bias the analysis model parameter estimates~\citep{hardtRainer:2012}.

The type of data social and behavioral scientists work with today often contains many variables.
For example, a single wave of the World Values Survey \citep{WVS:2020} contains more than 300 variables, and a single wave of the European Values Study \citep{EVS:2017} contains around 250 variables.
Running a MICE algorithm on this type of data, without selecting a subset of variables to use as predictors in the univariate imputation models, requires the algorithm to estimate regression models with hundreds of predictors for each variable under imputation.
With such a specification, the algorithm will be extremely slow, and the imputations will usually be poor.
However, selecting a smaller subset of predictors for each univariate imputation model can be a daunting task.
Choosing which predictors should be included in the univariate imputation models that constitute a run of the MICE algorithm entails a considerable degree of subjective judgment and requires both statistical and substantive expertise to achieve satisfactory results.

\citet[pp. 270--271]{vanBuuren:2018} provides a summary of different strategies an expert imputer can employ when designing imputation models for social science data sets with many variables.
The imputer can:
\begin{enumerate}
    \item \emph{Remove constants and collinear variables.}                      \label{item:colli}
        Collinear predictors will lead to unstable imputation model parameter estimates.
        Using one of a set of collinear predictors reduces the size of the predictor space for imputation models without losing any important information.
    \item \emph{Evaluate statistics describing the connection between variables in the data.} \label{item:iopat}
        For example, one can compute the proportion of usable cases for imputing a variable based on another \citep[i.e., inbound statistic][p. 108]{vanBuuren:2018}.
        The more cases that are usable, the more \textit{connected} the two variables are.
        The influx-outflux coefficients \citep[pp. 109--111]{vanBuuren:2018} provide overall measures of how each variable \textit{connects} to the rest of the data.
        In general, variables with high influx and outflux are preferred as predictors in an imputation model.
    \item \emph{Apply a correlation-thresholding strategy.}                 \label{item:quickpred}
        Only variables that are associated with the variables under imputation can be effective predictors in the imputation models.
        As a result, an intuitive strategy to select a small number of important predictors is to include only variables that correlate with the ones under imputation more strongly than a chosen threshold.
        However, the optimal threshold is not obvious. While choosing a low threshold might lead to selecting too many variables, choosing a high threshold might lead to excluding important predictors.
\end{enumerate}

These strategies are not guaranteed to avoid over-parameterization of the univariate imputation models, and more complex (combinations of) strategies are often needed.
The nature of some social science data sets offers a few other opportunities to reduce the dimensionality of the imputation models.
For example, with longitudinal data sets, the imputer might decide to use only the first measurement of the same construct when imputing other variables, or she may use the total score in place of the many items constituting a scale.
Additionally, an imputer can use high-dimensional prediction methods as univariate imputation models. 
Shrinkage methods, non-parametric prediction algorithms, and dimension reduction techniques can all be incorporated into MICE to reduce the complexity of the predictor selection step.

\citet{zhaoLong:2016} and \citet{dengEtAl:2016} proposed the use of lasso regression \citep{tibshirani:1996} within MICE.
Lasso is a shrinkage technique that can provide a data-driven selection of important predictors for each imputation model.
Decision trees are a popular class of semi-parametric prediction algorithms that can accommodate many predictor variables and represent complex, nonlinear relations among the variables~\citep{burgetteReiter:2010, dooveEtAl:2014, shahEtAl:2014}.
Decision trees have already been integrated into popular imputation software \citep[e.g., the R package \textit{mice},][]{mice}.
\citet{howardEtAl:2015} proposed using principal component analysis~\citep[PCA;][pp. 1--6]{jolliffe:2002} to reduce a set of auxiliary variables into a small set of principal components (PCs).
By extracting PCs from the (potentially numerous) auxiliary variables, this method can summarize the information contained in the auxiliary variables with just a few component scores.
These PCs can then be used as predictors in a standard, low-dimensional application of MICE\@.

The approaches described above have the potential to automatically address many of the issues caused by having too many variables available as potential imputation model predictors.
By default, high-dimensional prediction methods avoid collinearity issues and hence stabilize imputation model estimation.
Furthermore, many high-dimensional methods offer some form of variable selection or dimensionality reduction that can reduce the burden of making predictor choices.
Finally, algorithms that incorporate some form of regularization or dimension reduction allow the imputer to include more predictors in their imputation model, thereby increasing the chances of satisfying the MAR assumption.

A bespoke application of the MICE algorithm driven by subject-matter expertise may lead to better imputations than using automatic, data-driven approaches (though, as we illustrate in Section~\ref{sec:case-study}, this need not always be the case).
However, expert knowledge is not always available for every project that could benefit from imputation.
Furthermore, high-dimensional prediction models do not need to be the sole solution to the ``many variables'' imputation problem.
These methods can always be combined with expert knowledge to further improve the quality of imputation.

\cite{costantiniEtAl:2022a} compared a wide range of high-dimensional MI approaches, including the use of (Bayesian) lasso, ridge regression, random forests, correlation-thresholding, and PCA within the mice algorithm.
They found that using frequentist lasso to select the imputation model predictors and using PCA to reduce the dimensionality of the imputation models produced the best results.
The PCA-based approach was the strongest overall performer, though.
Incorporating PCA into the MICE algorithm consistently led to small estimation bias and close-to-nominal confidence interval coverage for the analysis model parameters.
However, that study considered only a single implementation of PCA\@ that was applied in a limited set of conditions.
In this paper, we extend the findings of \cite{costantiniEtAl:2022a} by further investigating the use of PCA within the MICE algorithm.
We use Monte Carlo simulation studies and a real-data case study to compare the performance of three alternative PCA-based MI approaches and evaluate how certain data characteristics impact that performance.
Two of these PCA-based MI approaches have been previously described in the literature.
We propose a novel third method here.

\subsection{Principal component analysis for MICE}\label{subsec:principal-component-analysis-for-mice}

PCA is a dimensionality reduction technique by which a set of variables is summarized with a smaller number of PCs.
These PCs are defined such that they explain the largest possible proportion of the original data's variance, given the number of PCs.
PCA can be used in conjunction with many statistical techniques, and its use in regression analysis has been extensive \citep[e.g.,][]{deJongKiers:1992, rosipalEtAl:2001, reissTodd:2007, parkEtAl:2021}.
In particular, one of the best-known uses of PCA in multiple regression is principal component regression~\citep[PCR;][pp. 168--173]{jolliffe:2002}, where PCs act as predictors in a multiple regression model.

Standard implementations of the MICE algorithm cycle through a sequence of univariate imputation models (i.e., one model for each incomplete variable).
Any, or all, of these univariate imputation models can be replaced by PCR\@.
We refer to this use of PCR in conjunction with MICE as MI-PCR\@.
The MI-PCR method is a broad approach that can be implemented in many different ways.
For example, a single set of PCs could be estimated before running MICE.
These PCs could then be used in a subsequent run of MICE as imputation model predictors.
Alternatively, a new PCA could be run within every iteration of MICE to produce updated PCs that incorporate the information from the most recent imputations.
These updated PCs could then be used as predictors to generate the imputations for that single iteration.
In this report, we primarily wish to investigate how different implementations of MI-PCR impact imputation quality.

We also explore how the number of components used as predictors in MI-PCR influences imputation quality.
Based on their analysis of data with a simple unidimensional latent structure, \citet{howardEtAl:2015} proposed retaining only the first PC\footnote{Since the first component explained 40\% of the variance in their simulations, \citet{howardEtAl:2015} alternatively recommended using the minimum number of components necessary to explain 40\% of the variance in the auxiliary variables.}.
However, medical and social science data are frequently characterized by complex latent structures that are unlikely to be well-summarized by a single component.
Therefore, evaluating the impact of the number of PCs is a secondary purpose of the present study.
Finally, as a tertiary focus, we also explore how key characteristics of the data affect MI-PCR.
In particular, we evaluate how the measurement level of the potential predictors and their strengths of association impact the performance of MI-PCR\@.

In this manuscript, we present the results of two Monte Carlo simulation studies through which we explore the performance of MI-PCR\@.
We assess this performance based on the estimation bias, confidence interval width, and confidence interval coverage (see Section \ref{subsubsec:analysis-and-outcome-measures} for details).
We also apply the different implementations of MI-PCR to the Fireworks Disaster data \citep[i.e., a real clinical psychology data set that has previously been used to demonstrate high-dimensional imputation problems,][p. 313]{vanBuuren:2018}.
In what follows, we describe how PCA can be used within the MICE algorithm (Section \ref{sec:mi-pcr-mice-using-pcr}).
We then discuss the simulation studies and the case study in Sections~\ref{sec:simulation-study},~\ref{sec:simulation-study-2}, and~\ref{sec:case-study}, respectively.
We provide a general discussion in Section~\ref{sec:discussion}, and share some final remarks on the selection of the number of components in Section~\ref{sec:final-remarks-npcs}.
We discuss limitations and future directions in Section~\ref{sec:limitations}.
Finally, we state our conclusions in Section~\ref{sec:conclusions}.
    
% Project:  paper-mipcr-compare
% Topic:    Description of imputation methods
% Author:   Edoardo Costantini
% Created:  2021-12-17
% Modified: 2022-12-23

\section{MI-PCR: MICE using PCR}\label{sec:mi-pcr-mice-using-pcr}

Here we briefly describe the MICE algorithm, PCA, and how PCA can be used in conjunction with MICE\@.
We use the following notation.
Scalars are denoted by lowercase letters in light typeface (non-bold).
Vectors and matrices are denoted by bold lowercase and bold uppercase letters, respectively.
We use the superscripts $obs$ and $mis$ to refer to the observed and missing elements in a vector.
For a given data set, we refer to the variables that are part of the researcher's model of scientific interest (e.g., the linear regression used to answer a research question) as the \text{analysis model variables}.
We refer to all other variables as \text{potential auxiliary variables} for the imputation models.
We use the subscripts $am$ and $av$ to refer to variables that are part of either the analysis model or the set of potential auxiliary variables, respectively.

\subsection{Multivariate imputation by chained equations}\label{subsec:multiple-imputation-chained-equations}

Consider an $n \times p$ data set $\bm{X}$.
Its columns, $\bm{x}_1, \dots, \bm{x}_p$, represent variables, and the rows represent observational units (e.g., people participating in a social survey).
Assume the first $t$ columns of $\bm{X}$ have missing values.
For each partially observed $\bm{x}_j$, with $j = 1, \dots, t$, the imputer defines a univariate imputation model:
\begin{equation} \label{eq:uim}
    f(\bm{x}_j|\bm{X}_{-j}, \bm{\theta}_j),
\end{equation}
where $\bm{X}_{-j}$ is the collection of variables in $\bm{X}$ excluding $\bm{x}_{j}$, and $\bm{\theta}_j$ is a vector of imputation model parameters.
Model \ref{eq:uim} is usually a generalized linear model chosen according to the measurement level of $\bm{x}_j$.
The MICE algorithm starts with replacing the missing values in each $\bm{x}_j$ with initial guesses.
Then, at every iteration, each variable is imputed by its univariate imputation model.
First, the imputation model parameters are drawn from their fully conditional posterior distributions, and then imputations are drawn from the posterior predictive distribution of $\bm{x}_j$.

For the $j$th variable under imputation at iteration $k$, the algorithm draws from the following distributions:
\begin{align}
    \bm{\theta}_j^{(k)} &\sim f(\bm{\theta}_j)f(\bm{x}_j^{obs}|\bm{X}_{-j}^{(k)}, \bm{\theta}_j) \label{eq:post_dist}, \\
    \bm{x}_j^{mis(k)} &\sim f(\bm{x}_j^{mis}|\bm{X}_{-j}^{(k)}, \bm{\theta}_j^{(k)})                \label{eq:pred_dist}
\end{align}
Equation~\ref{eq:post_dist} is the fully conditional posterior distribution defined as the product of $f(\bm{\theta}_j)$, a prior distribution for $\bm{\theta}_j$, and $f(\bm{x}_j^{obs}|\bm{X}_{-j}^{(k)}, \bm{\theta}_j)$, the likelihood of observing $\bm{x}_j^{obs}$ under the imputation model for $\bm{x}_j$.
Equation~\ref{eq:pred_dist} is the posterior predictive distribution from which updates of the imputations are drawn.
In both equations, $\bm{X}_{-j}^{(k)}$ is $(\bm{x}_1^{(k)}, \dots, \bm{x}_{j-1}^{(k)}, \bm{x}_{j+1}^{(k-1)}, \dots, \bm{x}_{p}^{(k-1)})$, meaning that at all times the most recently imputed values of all variables are used to impute other variables.

Each iteration comprises one complete cycle through all $t$ variables under imputation.
After a sufficient number of iterations, the algorithm converges to a stable equilibrium, and the imputations represent samples from the target multivariate distribution.
With this process, one can generate as many imputed data sets as desired.
Finally, the analysis model is estimated on each imputed data set, and the parameter estimates are pooled using Rubin's rules \citep{rubin:1987}.

\subsection{Principal component analysis}\label{subsec:principal-component-analysis}
PCA finds a low(er)-dimensional representation of $\bm{X}$ with minimal loss of information.
We refer to this low-dimensional representation as the $n \times q$ matrix $\bm{Z}$, where $q < p$.
The columns of $\bm{Z}$ are called the principal components (PCs) of $\bm{X}$.
The first PC of $\bm{X}$\footnote{We follow the common practice of assuming that the columns of $\bm{X}$ are mean-centered and scaled to have a variance of 1.} is the linear combination of the columns of $\bm{X}$ with the largest variance:
\begin{equation} \label{eq:z1}
    \bm{z}_1 = \bm{x}_1 w_{11} + \bm{x}_2 w_{12} + \dots + \bm{x}_p w_{1p} = \bm{X} \bm{w}_1,
\end{equation}
with $\bm{w}_1$ being the $p \times 1$ vector of weights $w_{11}, \dots, w_{1p}$.
The second principal component ($\bm{z}_2$) is defined by finding the vector of weights $\bm{w}_2$ giving the linear combination of $\bm{x}_1, \dots, \bm{x}_p$ with maximal variance out of all the linear combinations that are uncorrelated with $\bm{z}_1$.
Every subsequent column of $\bm{Z}$ can be understood in the same way: for example, $\bm{z}_3$ is the linear combination of $\bm{x}_1, \dots, \bm{x}_p$ that has maximal variance out of all the linear combinations that are uncorrelated with $\bm{z}_1$ and $\bm{z}_2$.
As a result, all PCs are uncorrelated by definition and every subsequent PC has a lower variance than the preceding one.
We can write the relationship between all the PCs and $\bm{X}$ in matrix notation:
\begin{equation} \label{eq:PCAmatnot}
    \bm{Z} = \bm{X} \bm{W},
\end{equation}
where $\bm{W}$ is a $p \times q$ matrix of weights, with columns $\bm{w}_1, \dots, \bm{w}_q$.
Equation~\ref{eq:PCAmatnot} also allows us to understand PCA as the process of projecting the original data from a $p$-dimensional space to a $q$-dimensional space.
The weight vectors $\bm{w}_1, \dots, \bm{w}_q$ define the directions in which the $n$ observations of $\bm{x}_1, \dots, \bm{x}_p$ are projected.
The projected values are the principal component scores $\bm{Z}$.

\subsection{Principal component regression}\label{subsec:principal-component-regression}

PCR replaces the $p$ predictors of a regression model with $q$ PCs extracted from those predictors.
Given the data $\bm{X}$, consider a standard regression model where the $j$th variable is regressed on the other columns in the data:
\begin{equation} \label{eq:lm}
    \bm{x}_j = \bm{X}_{-j}\bm{\beta} + \epsilon,
\end{equation}
where $\bm{x}_j$ is a $n \times 1$ vector of dependent variable scores, $\bm{\beta}$ is a $(p - 1) \times 1$ vector of $p - 1$ regression coefficients, and $\epsilon$ is a $n \times 1$ vector of independent normally distributed errors.
With PCR we use the PCs of $\bm{X}_{-j}$ in place of $\bm{X}_{-j}$ in the regression model so that Equation~\ref{eq:lm} can be rewritten as:
\begin{equation} \label{eq:PCR}
    \bm{x}_j = \bm{Z}\bm{\gamma} + \epsilon,
\end{equation}
where $\bm{\gamma}$ is a $q \times 1$ vector of regression coefficients.
The lower dimensionality of $\bm{Z}$ compared to $\bm{X}_{-j}$, and the independence of its columns, allow Equation~\ref{eq:PCR} to address the computational limitations of Equation~\ref{eq:lm} in the presence of many predictors.

\subsection{MI-PCR}\label{subsec:mi-pcr}
Standard MICE formulations are based on univariate imputation models that suffer from the same computational limitations that we discussed for multiple regression.
In general, the idea of MI-PCR is to use Equation~\ref{eq:PCR} as the univariate imputation model for every variable under imputation.
By doing so, we aim to address the ``many variables'' problem by summarising the imputation models' predictors with just a few PCs.
However, this idea can be implemented in many ways.
PCA can be used at different stages of the MICE algorithm, and different sets of variables can be summarized by the PCs.
In the following sections, we describe the three implementations evaluated in this study.

\subsubsection{MICE with PCA on auxiliary variables}

The most straightforward way to use PCA within MICE is to compute a single set of PCs based only on the potential auxiliary variables.
In general, potential auxiliary variables include predictors of missingness, variables related to the ones under imputation, and variables that are useless for imputation.
To reduce the dimensionality of imputation models, an expert imputer would usually examine these variables to locate and exclude members of the last group from the imputation models.
PCA can be used as an alternative, data-driven pre-processing step to project all the potential auxiliary variables onto a lower-dimensional space and bypass the need to select which variables to use as predictors in the imputation models.
We refer to this approach as MICE with PCA on the auxiliary variables (MI-PCR-AUX).

In MI-PCR-AUX, the univariate imputation models use as predictors the raw version of any variable that is part of the analysis model, and the principal components summarizing the potential auxiliary variables.
We can write the univariate imputation model as:
\begin{equation} \label{eq:MIPCRaux}
    f(\bm{x}_j|\bm{X}_{am, -j}, \bm{Z}_{av}, \bm{\theta}_j),
\end{equation}
where $\bm{X}_{am, -j}$ is the set of analysis model variables except the one under imputation, and $\bm{Z}_{av}$ is the set of PCs estimated from the set of potential auxiliary variables $\bm{X}_{av}$.
This use of PCA was proposed by~\cite{howardEtAl:2015}.

The strength of MI-PCR-AUX is using the raw analysis model variables (not filtered via the PCA) while including as much auxiliary information as possible (filtered via the PCA).
However, MI-PCR-AUX requires knowledge of the analysis model before imputation, and the possible presence of missing values in the potential auxiliary variables needs to be addressed.
\cite{howardEtAl:2015} suggested using single imputation to create a complete set of potential auxiliary variables, but implementing this idea is not necessarily straightforward.
All the obstacles to defining the univariate imputation models previously discussed still arise during this single imputation procedure.

\subsubsection{MICE with PCA on all variables}

One way to relax the requirements of knowing the analysis model before running the imputation procedure is to extract PCs from all available variables---including the ones under imputation---and then use only these PCs as predictors in the imputation models.
This approach (hereafter, MI-PCR-ALL) is implemented in the R package \textit{PcAux}~\citep{PcAux}.
The univariate imputation model for this approach can be written as:
\begin{equation} \label{eq:MIPCRall}
    f(\bm{x}_j|\bm{Z}, \bm{\theta}_j),
\end{equation}
where $\bm{Z}$ is the set of PCs estimated on $\bm{X}$.
Ideally, MI-PCR-ALL supports a wide range of analysis models, as the information on every available variable is summarized by the PCA procedure and included in all the imputation models.
In theory, the imputer could even augment $\bm{X}$ before extracting the PCs with every desired interaction and polynomial term that might be present in an analysis model.

As noted above, PCA cannot be performed in the presence of missing values.
Yet, to implement MI-PCR-ALL, we must perform PCA on all the variables in $\bm{X}$---even the ones targeted by imputation.
So, we must first (temporarily) treat the missing data to allow the PC extraction.
Our implementation of MI-PCR-ALL begins by filling in the missing values with a single imputation and extracting PCs from this completed data set.
It is important to note that these imputations will not be used for statistical inference.
So, the attenuated standard errors known to result from single imputation are not a concern.
As long as the imputations are well-constructed and consistent with the distribution of the original variables, inference based on data imputed with MI-PCR-ALL should not be negatively impacted.
Nevertheless, the performance of MI-PCR-ALL is tied to the quality of this first single imputation.

In MI-PCR-ALL, the PCs are the only predictors used in the univariate imputation models and do not have missing values.
Therefore, a single iteration of the MICE algorithm is sufficient.
Computationally, this is an advantage as there is no need to perform any burn-in iterations.

\subsubsection{MI with PCA on a variable-by-variable basis}

The most flexible way to incorporate PCA into MICE is to extract PCs at every iteration.
When imputing $\bm{x}_j$ at the $k$th iteration, PCs can be estimated from $\bm{X}^{(k)}_{-j}$ and used as predictors in the univariate imputation model.
Each univariate imputation model can then be defined as:
\begin{equation} \label{eq:MIPCR}
    f(\bm{x}_j|\bm{Z}_{-j}^{(k)}, \bm{\theta}_j),
\end{equation}
where $\bm{Z}_{-j}^{(k)}$ is the matrix storing the PC scores estimated on $\bm{X}^{(k)}_{-j}$.
We refer to this approach as MI-PCR-VBV because of the variable-by-variable use of PCA\@.

As with MI-PCR-ALL, MI-PCR-VBV does not require knowledge of the analysis model prior to imputation and it can support a wide range of analysis models.
Moreover, by extracting PCs at every iteration from variables with the most recently imputed values, MI-PCR-VBV addresses missing values in one step, without requiring a pre-processing single imputation procedure.
The disadvantage of MI-PCR-VBV is in the higher computational intensity relative to both MI-PCR-AUX and MI-PCR-ALL\@.
Performing PCA on large social surveys involves demanding matrix operations.
MI-PCR-VBV requires repeating these intensive manipulations for every variable under imputation and for every iteration of the MICE algorithm.
    
% Project:  paper-mipcr-compare
% Topic:    Methods for the paper
% Author:   Edoardo Costantini
% Created:  2021-12-17
% Modified: 2023-01-11

\section{Simulation study 1}\label{sec:simulation-study}

We investigated the relative performance of the methods described above with a Monte Carlo simulation study.
In particular, we were interested in assessing the estimation bias, confidence interval width, and confidence interval coverage of statistics estimated from the imputed data.
The purpose of this study was to evaluate these statistical properties of MI-PCR in several settings that differed in the proportion of noise variables present in the data, the measurement level of the variables, and the number of PCs used in the imputation models.

When defining the univariate imputation model, an expert imputer would usually exclude all variables that are weakly associated with the variables under imputation.
We refer to these weak predictors of the imputation targets as noise variables.
When using MI-PCR, noise variables will contribute to the construction of the PCs as much as the important predictors of the variables under imputation.
Consequently, PCs extracted from data that contain a large proportion of noise variables may be more weakly associated with the variables under imputation.
We expect that the presence of a larger proportion of noise variables will negatively impact the performance of MI-PCR\@ (i.e., larger bias, lower efficiency, larger deviation from nominal coverage).
Additionally, in real survey applications, theoretical constructs of interest are often measured with discrete items such as Likert scales.
The number of categories with which information is recorded in a variable can impact on how well $\bm{Z}$ represents $\bm{X}$.
Finally, each of the implementations of MI-PCR described above can be used with different numbers of PCs.
\cite{howardEtAl:2015} suggested that using the first PC may be sufficient.
However, they used a set of strongly associated potential auxiliary variables measuring a single latent factor.
When the underlying correlation structure is more complex (i.e., more than one latent factor, different correlation levels) using only the first PC is likely to result in a poor representation of the data and poor imputation performance.
In what follows we outline the simulation study procedure, discuss the experimental factors in detail, and report the results.

\subsection{Method}\label{subsec:method}

The simulation study procedure involved five steps:
\begin{enumerate}
    \item Data generation: We simulated $S = 500$ data sets from a confirmatory factor analysis model, following the procedure described in Section \ref{subsubsec:data-generation}.
    \item Missing data imposition: We imposed missing values on four target items in each generated data set, following the procedure described in Section \ref{subsubsec:missing-data-imposition}.
    \item Imputation: For each incomplete data set, we applied each of the different imputation methods to generate $d$ multiply imputed data sets, as described in Section \ref{subsubsec:imputation-procedures}.
    \item Analysis: We used the $d$ imputed data sets to estimate the means, variances, covariances, and correlations of the four items with missing values, and we pooled the estimates according to Rubin's rules \citep[p. 76]{rubin:1986}.
    \item Evaluation: We assessed the performance of each imputation method by computing the bias, confidence interval width, and confidence interval coverage of the above statistics, as described in Section \ref{subsubsec:analysis-and-outcome-measures}.
\end{enumerate}

\subsubsection{Data generation}\label{subsubsec:data-generation}

For each replication, we generated a $500 \times 56$ matrix of fully observed data $\bm{X}$.
We fixed the sample size to 500 observations to generate data sets that would have statistical properties similar to large social science data sets without needlessly increasing the computational demands of the simulation study.
Each data set was generated based on the following confirmatory factor analysis model:
\begin{equation} \label{eq:CFAmod}
    \bm{X} = \bm{F}\bm{\Lambda}' + \bm{E},
\end{equation}
where $\bm{F}$ is a $500 \times 7$ matrix of latent variables scores, $\bm{\Lambda}$ is a $56 \times 7$ matrix of factor loadings, $\bm{\Lambda}'$ is its transpose, and $\bm{E}$ is a $500 \times 56$ matrix of measurement errors.
The dimensionality of the data resembles that of short-scale questionnaires used in the social sciences.
For example, consider the NEO Five-Factor Inventory \citep[][NEO-FFI]{costaMcCrae:2008}, which measures the Big Five personality (i.e., Extraversion, Agreeableness, Conscientiousness, Emotional Stability/Neuroticism, and Openness to Experience) with 12 items each, for a total of $5 \times 12 = 60$ items.

The factor loading matrix $\bm{\Lambda}$ described a simple measurement structure~\citep[i.e., a structure wherein every item loads on a single factor,][p. 234]{bollen:1989}.
The factor loadings were set to the fixed value of $\lambda = 0.85$ to represent a plausible, but reasonably high, item-scale association.
We generated data with relatively high factor loadings because we wanted to mitigate the impact of measurement error on our findings without resorting to implausibly precise data.
We sampled the latent scores for seven factors from a multivariate normal distribution with mean $\bm{0}$ and covariance matrix $\bm{\Psi}$:
\begin{equation} \label{eq:Psi}
\bm{\Psi} =
    \begin{bmatrix}
    1           & \psi_{12} & \dots & \psi_{17} \\
    \psi_{21}   & 1         & \dots & \psi_{27} \\
    \dots       & \dots     & \dots & \dots \\
    \psi_{71}   & \psi_{72} & \dots & 1 \\
    \end{bmatrix}.
\end{equation}
The matrix of measurement errors $\bm{E}$ was sampled from a multivariate normal distribution with mean $\bm{0}$ and covariance matrix $\bm{\Theta}$.
The off-diagonal elements of $\bm{\Theta}$ were set to 0 to reflect uncorrelated errors, while the diagonal elements were specified as $1 - \lambda^2$ to give the simulated items unit variances.
After sampling, data were rescaled to have approximately a mean of 5 and a variance of 6.5, which are common values for Likert items in social surveys measured on a 10-point scale.

Each data matrix generated with the procedure described above was partitioned into three sub-matrices:
\begin{equation}
    \bm{X} = (\bm{T}, \bm{M}, \bm{A}) \label{eq:Xpart}
\end{equation}
where:
\begin{itemize}
    \item $\bm{T}$ is an $n \times 4$ matrix consisting of the first four indicators ($\bm{x}_1, \dots, \bm{x}_4$) of the first latent variable $\bm{f}_1$. 
    We imposed missing data on these items as described in Section \ref{subsubsec:missing-data-imposition}.
    \item $\bm{M}$ was an $n \times 4$ matrix consisting of the other four items ($\bm{x}_5, \dots, \bm{x}_8$) measuring the first latent variable.
    These items were used to define the probability of nonresponse for the items in $\bm{T}$ as described in Section \ref{subsubsec:missing-data-imposition}.
    \item $\bm{A}$ was an $n \times 48$ matrix consisting of 48 items measuring the the remaining six latent variables $\bm{f}_2, \dots, \bm{f}_7$.
\end{itemize}

In generating the data, we varied two design factors: the number of categories into which the potential auxiliary variables were coarsened ($nCat = \infty, 7, 5, 3, 2$), and the proportion of noise variables ($pn$ = 0, 0.33, 0.67, 1).
We crossed these factors in a $5(nCat) \times 4(pn)$ factorial design.
The variables in $\bm{M}$ and $\bm{A}$ represented the pool of potential auxiliary variables.
We discretized these variables according to the $nCat$ factor to study the impact of data coarseness on the performance of the imputation methods.
The $nCat = \infty$ level represents the uncoarsened, continuous variables.
Although the data were coarsened, we applied the PCA underlying the MI-PCR methods to the Pearson correlation matrix.
We recognize that we could have used polychoric, polyserial, or tetrachoric correlations, yet we purposefully chose not to do so.
In the context of imputation, using alternative correlation computations impacts the imputations only through the predictive performance of PCR.
\citet{kolenikovAngeles:2009} showed that estimating PCA based on the polychoric correlation instead of the Pearson correlation did not improve the predictive performance of PCR when applied to reduce the dimensionality of a set of ordinal predictors.
Furthermore, extracting PCs based on these alternative correlations is more computationally expensive than using Pearson correlations.
Considering the lack of expected advantages and the higher computational load of these alternatives, we decided to treat the ordinal data as numeric and estimate the PCA from Pearson correlations.

We used the $pn$ factor to define the proportion of noise variables in $\bm{A}$.
That is, the proportion of items in $\bm{A}$ that are uncorrelated with the items in $\bm{T}$.
We controlled this factor at the latent variable level through the values of $\psi_{jk}$, the latent correlation in Equation~\ref{eq:Psi}.
When $pn = 0.33$, two out of the six latent variables indicated by the items in $\bm{A}$ had a low correlation ($0.1$) with $\bm{f}_1$, and the remaining four had a high correlation ($0.7$) with $\bm{f}_1$.
As a result, one-third of the items in $\bm{A}$ were also lowly correlated with the items in $\bm{T}$ and $\bm{M}$.
When $pn = 0$, all latent variables were correlated at 0.7, so every variable in $\bm{A}$ correlated highly with the variables in $\bm{T}$ and $\bm{M}$.
When $pn = 1$, all latent variables were correlated at 0.1, so all variables in $\bm{A}$ were trivially correlated with the variables in $\bm{T}$ and $\bm{M}$.

\subsubsection{Missing data imposition} \label{subsubsec:missing-data-imposition}

We imposed missing values in $\bm{T}$ by first generating an indicator of missingness ($\delta$) for each column of $\bm{T}$.
When the indicator took value 0, we left the original sampled value; when the indicator took value 1, we replaced the sampled value with a missing value.
The indicator was produced by sampling from Bernoulli distributions with probabilities defined based on the following logit model:
	\begin{equation} \label{eq:logit}
		logit(\delta = 1) = \beta_0 + \bm{M}\bm{\beta},
	\end{equation}
where $\beta_0$ is an intercept parameter, and $\bm{\beta}$ is a vector of slope parameters.
Because only the variables in $\bm{M}$ were used to predict missingness in $\bm{T}$, the probability of nonresponse for a variable never depended on the variable itself.
We defined the value of $\beta_0$ to align the missing values with the positive tail of $\bm{M}\bm{\beta}$, which is a mechanism known as \emph{right-tail MAR} \citep{schoutenVink:2021}.

All slopes in $\bm{\beta}$ were fixed to 1, while the value of $\beta_0$ was chosen with an optimization algorithm that minimized the difference between the actual and desired proportion of missing values\footnote{
    The pseudo R-squared for the logistic regression of the missing value indicator on the predictors of missingness was approximately 14\%.
    The AUC for the logistic regression was approximately 0.74.
    }.
We fixed the proportion of missing values for each variable to 0.3, which represents a realistic---but reasonably large---value for social science data \citep{vink:2016}.
We selected this proportion of missing cases to be plausible for social science data yet large enough to create substantial problems if the missing values were poorly treated\footnote{We report plots showing the impact of the response mechanism we used on complete case analysis in the supplementary material.}.

Missingness was imposed using $\bm{M}$ in its original continuous form, even in those conditions where the potential auxiliary variables were coarsened (i.e., $nCat \neq \infty$).
We made this decision to maintain the strength of the MAR mechanism as consistently as possible across conditions.
Using the coarsened versions of $\bm{M}$ to impose missing values would have generated a weaker MAR relation (closer to missing completely at random, MCAR) for conditions with lower numbers of categories.
At the same time, the solution we adopted is also imperfect.
For a lower number of categories, the data available to use as predictors in the imputation models were worse representations of the actual MAR predictors.
As a result, one might argue that, for the conditions with fewer categories, imputation was closer to a missing not at random (MNAR) situation rather than MAR.
When designing the study we reasoned that making imputation more difficult (closer to MNAR) rather than easier (closer to MCAR) would lead to more informative results.

\subsubsection{Imputation procedures}\label{subsubsec:imputation-procedures}

After generating the data and imposing missing values, every variable in $\bm{T}$ was imputed with the three versions of MI-PCR described above:
 \begin{itemize}
    \item MI-PCR-AUX\@.
            The PCs used in the univariate imputation model for the $j$th variable were estimated from the set of potential auxiliary variables (the variables in $\bm{M}$ and $\bm{A}$).
            For every variable under imputation, the other variables in $\bm{T}$ were also used as predictors.
    \item MI-PCR-ALL\@.
            The PCs used in the univariate imputation model for the $j$th variable were estimated from the entire data set ($\bm{T}, \bm{M}, \bm{A}$).
            An initial single imputation step was required to obtain a complete version of $\bm{T}$ from which estimate the PCs.
            We implemented this imputation by running a single chain of the mice algorithm for 20 iterations.
            We selected the predictors for this single imputation model via a correlation-thresholding strategy whereby all variables correlating at least $r = 0.3$ with the imputation targets were used as predictors.
    \item MI-PCR-VBV\@.
            The PCs used in the univariate imputation model for the $j$th variable were estimated from all other variables ($\bm{T}_{-j}, \bm{M}, \bm{A}$), at every iteration.
\end{itemize}

We also imputed the missing data using two non-PCR methods to act as  additional points of comparison:
\begin{itemize}	
\item MI with correlation-based thresholding (MI-QP)\@.
            As a pragmatic point of comparison, this method used the \emph{quickpred} function from the R package mice \citep{mice} to select the predictors for the univariate imputation models via the correlation-based thresholding strategy described by \citet[pp. 687-–688]{vanBuurenEtAl:1999}.
            To implement this approach, we selected only those predictors that correlated with the imputation targets (or their associated missingness indicators) at $r = 0.1$ or higher.
            For every $j$th variable under imputation, \emph{quickpred} selected predictors from the remaining variables ($\bm{T}_{-j}, \bm{M}, \bm{A}$).
    \item MI with oracle properties (MI-OR)\@.
            This method represented the idealized, hypothetical situation wherein the imputer knows the optimal imputation model.
	    The univariate imputation models included the remaining analysis model variables (which were also the other imputation targets, in this case) and the predictors that were used to impose missingness ($\bm{T}_{-j}, \bm{M}$).
            For this method, we used our perfect knowledge of the missing data mechanism to define which variables should be predictors in the imputation models.
            As such, MI-OR represents an optimal point of comparison but is not replicable in practice.
\end{itemize}

To explore how the number of PCs used in MI-PCR impacts performance, we implemented the MI-PCR methods with different numbers of components.
We implemented each method with fixed numbers of components (i.e., 1, 2, $\dots$, 10, 20, 25) as well as the maximum number of components possible (which varied by method).
In PCA, the maximum number of components cannot exceed the number of rows or columns of the data, so this number depends on the specific MI-PCR implementation:
\begin{itemize}
\item For MI-PCR-AUX, the maximum number of PCs was $56 - 4 = 52$, the number of variables in matrices $\bm{M}$ and $\bm{A}$.
\item For MI-PCR-ALL, the maximum was $56$, the total number of variables in the data set.
\item For MI-PCR-VBV, the maximum was $56 - 1 = 55$, the number of variables available as predictors for each univariate imputation model.
\end{itemize}
Using the maximum number of components addresses possible collinearity among the imputation model predictors without performing any dimensionality reduction.

Every imputation algorithm was used to obtain five imputed data sets, the default in the \emph{mice} R package.
All starting imputations were created by a simple random draw from the data.
We set the number of iterations to 20 after checking convergence for a subset of replications.
We evaluated convergence by plotting the means and standard deviations of the imputed variables.
For more information on this approach, see \citet{costantiniEtAl:2022a}.
Convergence plots are provided in the supplementary material\footnote{The interested reader can interact with the trace plots through a Shiny app that can be downloaded and installed as an R package \citep{costantini:2022a}}.

\subsubsection{Analysis and outcome measures} \label{subsubsec:analysis-and-outcome-measures}

For each of the $S = 500$ simulated data sets, we imputed the variables in $\bm{T}$ with the different methods described above, and we pooled the estimates of their means, variances, covariances, and correlations\footnote{We applied Fisher's $z$ transformation to the correlation coefficients before pooling.
We then back-transformed the pooled correlation coefficient estimates with the inverse Fisher's $z$ transformation \citep[p. 146]{vanBuuren:2018}.} across the multiple imputations.
The pooled estimates were stored and used to assess the performance of the imputation methods.
For a given parameter $\phi$ (e.g., mean of $\bm{x}_1$, correlation between $\bm{x}_1$ and $\bm{x}_2$), we used the absolute percent relative bias (PRB) to quantify the estimation bias introduced by the imputation procedure:
\begin{equation} \label{eq:prb}
\textit{PRB} = \displaystyle\left\lvert\ \frac{\bar{\hat{\phi}} - \phi}{\phi} \right\rvert \times 100
\end{equation}
where $\phi$ is the true value of the focal parameter defined as
$\sum_{s=1}^{S} \hat{\phi}_{s}^{full}/S$
, with
$\phi_{s}^{full}$
being the parameter estimate for the $s$th repetition computed on the original fully observed data.
The averaged focal parameter estimate under a given missing data treatment was computed as
$\bar{\hat{\phi}} = \sum_{s=1}^{S} \hat{\phi}_{s}/S$,
with
$\hat{\phi}_{s}$ being the estimate obtained from the treated incomplete data in the
$s$th simulated data set.
Following~\cite{muthenEtAl:1987}, we considered $\text{PRB} > 10$ as indicative of problematic
estimation bias.

To measure the statistical efficiency of the imputation methods we computed the average width of
the confidence intervals (CIW).
\begin{equation} \label{eq:ciw}
CIW = \frac{\sum_{s=1}^{S} \widehat{CI}^{upper}_{s} - \widehat{CI}^{lower}_{s}}{S},
\end{equation}
with $\widehat{CI}^{upper}_{s}$ and $\widehat{CI}^{lower}_{s}$ being the upper and lower bounds of the estimated confidence interval for the $s$th repetition.
In general, narrower CIWs indicate higher efficiency.
However, narrower CIWs are not preferred if they come at the expense of good confidence interval coverage (CIC) of the true parameter values.
CIC is the proportion of confidence intervals that contain the true value of the parameter, across the $S$ simulated data sets:
\begin{equation} \label{eq:cic}
CIC =  \frac{ \sum_{s=1}^{S} I(\phi \in \widehat{CI}_s ) }{S},
\end{equation}
where $\widehat{CI}_s$ is the confidence interval of the parameter estimate $\hat{\phi}_{s}$ in the $s$th replication, and $I(.)$ is the indicator function that returns 1 if the argument is true and 0 otherwise.
CIC depends on both the bias and the CIW for a parameter estimate.
An imputation method with good coverage should result in CICs greater than or equal to the nominal rate.
For 95\% CIs, CIC below 0.9 is usually considered problematic (e.g., \citealp[p. 52]{vanBuuren:2018}; \citealp[p. 340]{collinsEtAl:2001}) as it implies inflated Type I error rates.
High CIC (e.g., 0.99) implies inflated Type II error rates.

\subsection{Results}\label{subsec:results}

\subsubsection{Bias}

\begin{figure}
\centering
\begin{knitrout}
\definecolor{shadecolor}{rgb}{0.969, 0.969, 0.969}\color{fgcolor}

{\centering \includegraphics[width=\maxwidth]{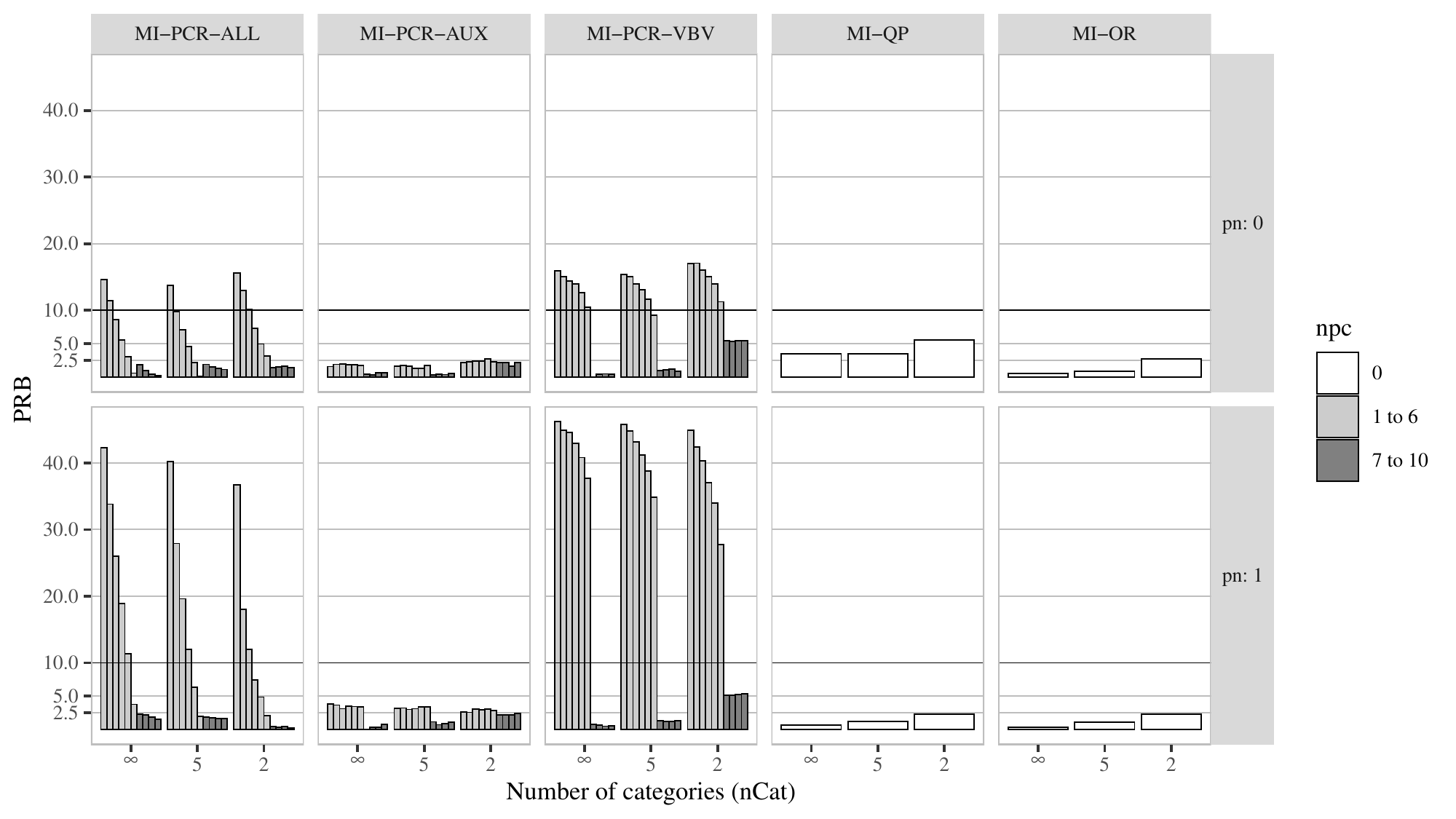} 

}

\end{knitrout}
\caption{\label{fig:bias}   
	Percent relative bias for the correlation between $\bm{x}_1$ and $\bm{x}_2$ in simulation study 1.
    $pn$ is the proportion of noise variables in $\bm{A}$.
    $npc$ is the number of PCs used by a given imputation method.
    The X-axis of each histogram distinguishes three levels of coarsening for the potential auxiliary variables ($nCat = (\infty, 5, 2)$).
    For each MI-PCR method, we reported a different vertical bar for each PRB obtained using a different number of PCs (from 1 to 10, from left to right).
	}
\end{figure}

In Figure~\ref{fig:bias}, we report the PRB for the correlation coefficient between $\bm{x}_1$ and $\bm{x}_2$---two of the four imputed items in $\bm{T}$---in an illustrative selection of conditions.
In this report, we focus on the estimates of the correlation as this was the hardest parameter to recover (i.e., the performance differences were most pronounced).
In the supplementary material\footnote{The interested reader can choose which results to plot with an interactive Shiny app that can be downloaded and installed as an R package \citep{costantini:2022a}}, we report the same figures for the mean, variance, and covariance.
In what follows, we write the number of components used for PCR-based methods in subscript, so that MI-PCR-AUX$_{(1)}$ refers to the use of MI-PCR-AUX with a single component.
Similarly, we use the subscript to discuss the performance of PCR-based methods using a range of PCs.
For example, we refer to the performance of MI-PCR-VBV using 7 to 10 PCs as MI-PCR-VBV$_{\text{(7:10)}}$.

MI-PCR-AUX resulted in acceptable bias in all conditions (PRB $< 10$.)
The bias resulting from MI-PCR-AUX depended on the number of PCs retained as predictors.
The bias obtained by MI-PCR-AUX$_{\text{(1:6)}}$ was around a PRB of 2.5 for all the levels of $nCat$ and $pn$.
MI-PCR-AUX$_{\text{(7:10)}}$ resulted in PRBs below 2.5 for $nCat = \infty$ and $5$, while the bias increased to approximately 2.5 for $nCat = 2$, for both $pn = 0$ and $1$.

Independently of the coarseness of the data, MI-PCR-VBV$_{\text{(1:6)}}$ resulted in PRBs between 10 and 20, and above 20, for $pn = 0$ and $1$, respectively.
MI-PCR-VBV$_{\text{(7:10)}}$ led to PRBs below 2.5 for $nCat = \infty$ and $5$, and to PRBs around 5 for $nCat = 2$.
These values were not affected by the varying proportion of noise variables.

In the condition with no noise variables, MI-PCR-ALL$_\text{(3)}$ already returned PRB smaller than 10 for $nCat = \infty$ and $5$, while for $nCat = 2$ MI-PCR-ALL needed at least 4 components to return acceptable bias.
In the conditions with $pn = 1$, the number of components needed by MI-PCR-ALL to produce PRB $< 10$ were 6, 5, and 4, for $nCat$ $\infty$, 5, and 2, respectively.
As with the other MI-PCR methods, MI-PCR-ALL$_{\text{(7:10)}}$ resulted in low bias (PRB $< 2.5$) for both $pn = 0$ and $1$.
With $pn = 1$ and $nCat = 2$, MI-PCR-ALL$_{\text{(7:10)}}$ resulted in lower bias compared to other levels of $nCat$, and in the smallest bias across all methods.

MI-QP resulted in acceptable bias in all conditions (PRB $< 10$.)
The PRB obtained with this method increased as a function of the coarseness of the data: the smallest PRB was obtained for $nCat = \infty$ and the highest was obtained for $nCat = 2$.
Furthermore, the bias obtained by MI-QP was smaller for $pn = 1$ than for $pn = 0$.
Finally, MI-OR produced PRBs below 2.5 in all conditions and, while this performance was not affected by the proportion of noise variables, the bias slightly increased as the data were coarsened to fewer categories.

% Read all plots for the low dimensional condition

% Plot PRB

\subsubsection{Confidence Intervals}

\begin{figure}
\centering
\begin{knitrout}
\definecolor{shadecolor}{rgb}{0.969, 0.969, 0.969}\color{fgcolor}

{\centering \includegraphics[width=\maxwidth]{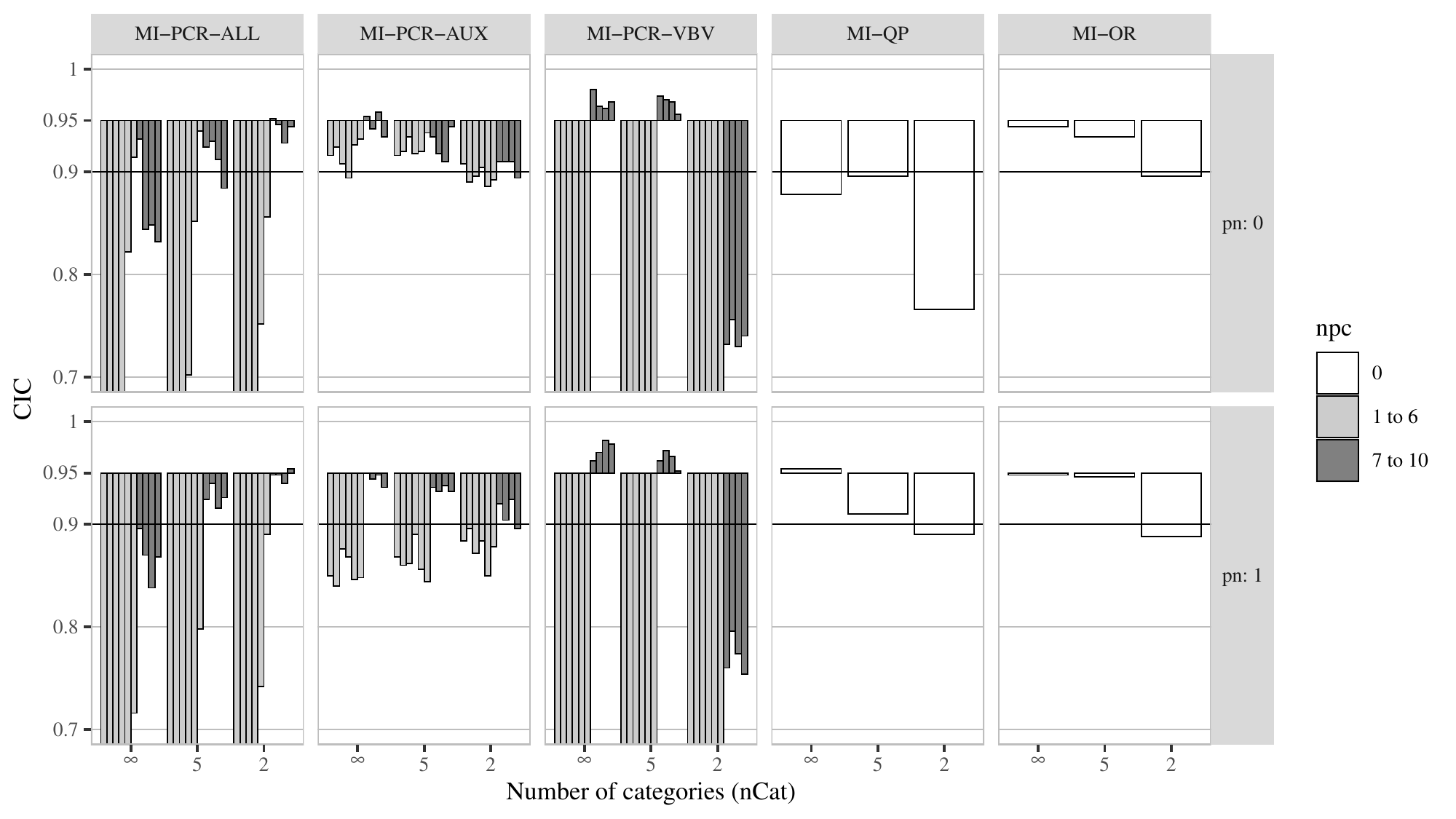} 

}

\end{knitrout}
\caption{\label{fig:cic}
	Confidence interval coverage for the correlation between $\bm{x}_1$ and $\bm{x}_2$ in simulation study 1.
    $pn$ is the proportion of noise variables in $\bm{A}$.
    $npc$ is the number of PCs used by a given imputation method.
    The X-axis of each histogram distinguishes three levels of coarsening for the potential auxiliary variables ($nCat = (\infty, 5, 2)$).
    For each MI-PCR method, we reported a different vertical bar for each CIC obtained using a different number of PCs (from 1 to 10, from left to right).
	}
\end{figure}

CICs for the correlation coefficient between $\bm{x}_1$ and $\bm{x}_2$ are plotted in Figure~\ref{fig:cic}.
As a general trend, the fewer categories used for discretization, the higher the deviation from nominal coverage was.
MI-PCR-ALL was the only exception to this trend, showing lower deviations of CIC from 0.95 for lower numbers of categories.

MI-PCR-AUX$_{\text{(1:6)}}$ resulted in small deviations from nominal coverage (CIC $\approx 0.9$) for $pn = 0$ and in clear under-coverage (CIC $< 0.9$) for $pn = 1$.
These trends were constant across the different levels of $nCat$.
MI-PCR-AUX$_{\text{(7:10)}}$ resulted in acceptable coverage (CIC between $0.9$ and $0.95$) independently of $pn$, but smaller numbers of categories led to more deviation from nominal coverage.
MI-PCR-VBV$_{\text{(1:6)}}$ led to severe under-coverage (CIC $< 0.7$) in all data conditions.
MI-PCR-VBV$_{\text{(7:10)}}$ resulted in close to nominal coverage, with a tendency toward over-coverage (CIC $> 0.95$), in all conditions except the ones with $nCat = 2$, when it resulted in severe under-coverage (CIC $< 0.8$), for both $pn = 0$ and $1$.
MI-PCR-ALL$_{\text{(1:6)}}$ led to CIC $< 0.9$ in all data conditions, expect for MI-PCR-ALL$_{(6)}$ which produced CIC between $0.9$ and $0.95$ for $pn = 0$ and $nCat = \infty$ and $5$.
MI-PCR-ALL$_{\text{(7:10)}}$ showed more severe under-coverage than the other MI-PCR methods in all conditions except for $nCat = 2$, for both $pn = 0$ or $1$.
Finally, MI-QP resulted in approximately nominal coverage only in the condition with $pn = 1$ and $nCat = \infty$, and MI-OR resulted in under-coverage only for $nCat = 2$.

% Plot CIC

\begin{figure}
\centering
\begin{knitrout}
\definecolor{shadecolor}{rgb}{0.969, 0.969, 0.969}\color{fgcolor}

{\centering \includegraphics[width=\maxwidth]{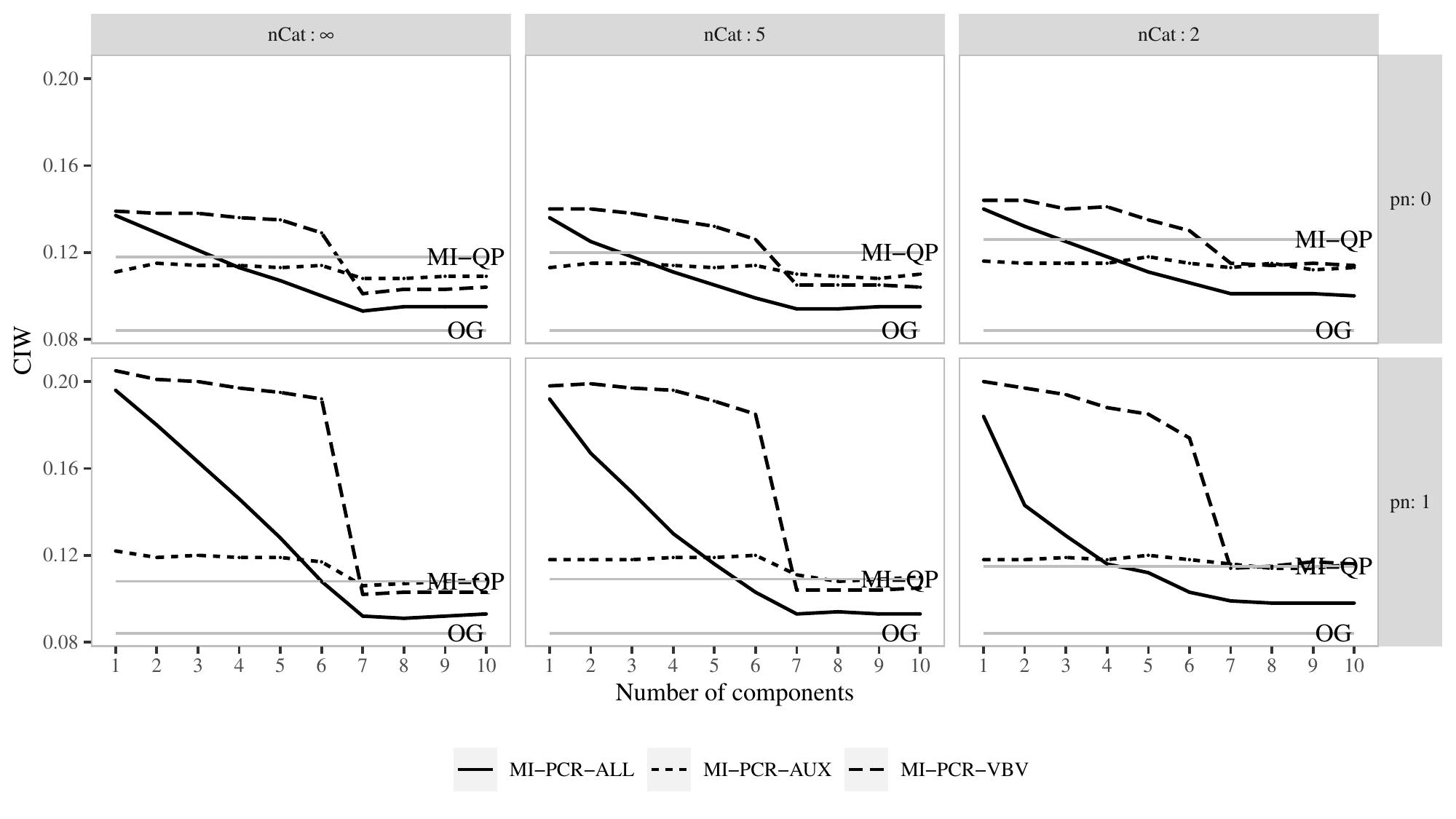} 

}

\end{knitrout}
\caption{\label{fig:ciw}
	Average confidence interval width for the correlation between $\bm{x}_1$ and $\bm{x}_2$ in simulation study 1.
    $nCat$ is the number of categories for the items in matrices $\bm{M}$ and $\bm{A}$.
    $pn$ is the proportion of noise variables in $\bm{A}$.
	}
\end{figure}

Figure~\ref{fig:ciw} shows the average CIW for the correlation between $\bm{x}_1$ and $\bm{x}_2$.
All MI-PCR methods using at least seven components produced narrower confidence intervals than the intervals obtained with MI-QP\@.
MI-PCR-ALL$_{\text{(7:10)}}$ resulted in the narrowest confidence intervals, followed by MI-PCR-VBV$_{\text{(7:10)}}$, and MI-PCR-AUX$_{\text{(7:10)}}$\@.
However, the confidence intervals obtained with MI-PCR-ALL and MI-PCR-VBV almost doubled in size when using fewer than seven components.
MI-PCR-AUX$_{\text{(1:6)}}$ was less influenced by the number of PCs, providing only slightly wider confidence intervals than MI-PCR-AUX$_{\text{(7:10)}}$\@.
MI-PCR-VBV$_{\text{(7:10)}}$ and MI-PCR-AUX$_{\text{(7:10)}}$ resulted in approximately the same CIW independently of the coarseness of the data and the proportion of noise variables.
For $pn = 1$, MI-PCR-ALL$_{\text{(1:6)}}$ resulted in narrower confidence intervals when data were dichotomized compared to when they were not, while MI-PCR-ALL$_{\text{(7:10)}}$ resulted in approximately the same CIW, independently of the data coarseness.

\subsection{Discussion}

The most important factor influencing the performance of the MI-PCR methods was the number of components used.
This result followed a very clear, dichotomous pattern: using fewer than seven components led to poor performance across all outcome measures for all MI-PCR methods, whereas using more than seven components universally produced much better performance.
As we generated the data based on seven latent variables, this result suggests that researchers using MI-PCR must employ at least as many components as there are latent variables in the data-generating model.
Thankfully, except for MI-PCR-ALL, our results suggest no substantial consequences for using more components than necessary.

For all methods except MI-PCR-ALL, bias was higher when the potential auxiliary variables were discretized to fewer categories.
This pattern probably reflects the loss of observed information caused by discretization.
As a result of discretization, the association between variables is attenuated, which makes every auxiliary variable less useful as an imputation model predictor.
Furthermore, missing values were imposed on the target variables based on the continuous variables in set $\bm{M}$.
When the variables in set $\bm{M}$ were discretized, they became poorer representations of the actual MAR predictors.
As a result, even MI-OR follows a trend of increasing bias for decreasing numbers of categories.
The discretization of the potential auxiliary variables did not seem to impact the performance of MI-PCR more than other approaches.

MI-QP was strongly influenced by the proportion of noise variables in the set of potential auxiliary variables.
The \textit{quickpred} approach is most effective when the proportion of noise variable is high, because there is a clear distinction between variables that are correlated with the targets of imputation and those that are not.
However, the method loses its efficacy as more variables correlate strongly with the imputation targets, because a large number of nearly collinear predictors end up selected into the model.
For the most part, the proportion of noise variables did not have a strong impact on the performance of the MI-PCR methods.
A higher proportion of noisy variables resulted in higher bias when fewer than seven components were selected.
However, when enough PCs were retained, the MI-PCR methods' performances were indistinguishable across different proportions of noise variables.

The variable-by-variable approach seems to be the most promising way of using PCA within MICE.
Although sometimes outperformed by MI-PCR-AUX (e.g., when using fewer than seven components or when the potential auxiliary variables were dichotomized), MI-PCR-VBV produced low bias, good coverage, and its competitive imputation performance was also accompanied by a few other desirable features.
Compared to the other MI-PCR approaches, when MI-PCR-VBV deviated from nominal coverage, it showed a tendency toward over-coverage.
An imputation method characterized by over-coverage will inflate type II error rates, making inferences more conservative than they should be.
Although this is undesirable, it is usually less worrisome than the problem of under-coverage, which inflates type I error rates.
Furthermore, MI-PCR-VBV does not rely on an initial single imputation step to obtain complete data for extracting PCs.
By performing PCA at every iteration, there is no need to pre-impute the variables from which PCs are extracted.
Finally, MI-PCR-VBV does not require knowledge of the analysis model, while MI-PCR-AUX needs this knowledge to distinguish which variables in the data should be summarized by PCs and which variables should be used in their raw form.
At the same time, MI-PCR-VBV can still incorporate important features of the analysis model, if these features are known before imputation.
Analysis model variables can be included in any desired functional form as predictors in the imputation models and excluded from the PC estimation.
In such a scenario, MI-PCR-VBV would supplement each imputation model with PCs representing information that would have otherwise been ignored by the imputation procedure.

\section{Simulation study 2: More variables}\label{sec:simulation-study-2}

Psychological self-report inventories and large social surveys can have hundreds of variables.
For example, the NEO-PR-I \citep{costaEtAl:1991} measures the same 5 personality factors as the NEO-FFI but uses 48 (instead of 12) items to define each factor.
Consequently, this single personality inventory comprises 240 items.
To evaluate the extent to which the results from the above simulation study generalize to problems with more variables, we replicated simulation study 1 with a larger number of potential auxiliary variables.
A larger pool of potential auxiliary variables could cause problems in a couple of ways.
If more auxiliary variables bring a larger number of important imputation predictors, there is an increased risk of collinearity among these predictors.
Likewise, if more auxiliary variables produce more noise variables, the added noise could reduce the effectiveness of PCA as a dimensionality reduction technique (at least, with respect to the task of generating imputation model predictors).

In the second simulation study, we were only interested in exploring whether the relative performance of the methods studied was impacted by a larger dimensionality of the auxiliary set.
We did not want to confound the comparison by altering the nature of the missing data problem, as well.
Hence, we increased the size of the auxiliary set, $\bm{A}$, by increasing the number of items measuring the latent variables in $\bm{A}$ from 8 to 39.
We kept the number of items measuring the first latent variable equal to 8 to keep the missing data problem comparable between the two simulation studies (i.e., same number of items under imputation, same number of MAR predictors, same correlation between variables with missing values and MAR predictors).
As a result, the data sets we generated for the second simulation study comprised $8 + 6 \times 39 = 242$ variables.
Otherwise, we used the same simulation study procedure described in~\ref{subsec:method} to generate the data, impose missing values, perform imputations, and analyze the results.

\begin{figure}
\centering
\begin{knitrout}
\definecolor{shadecolor}{rgb}{0.969, 0.969, 0.969}\color{fgcolor}

{\centering \includegraphics[width=\maxwidth]{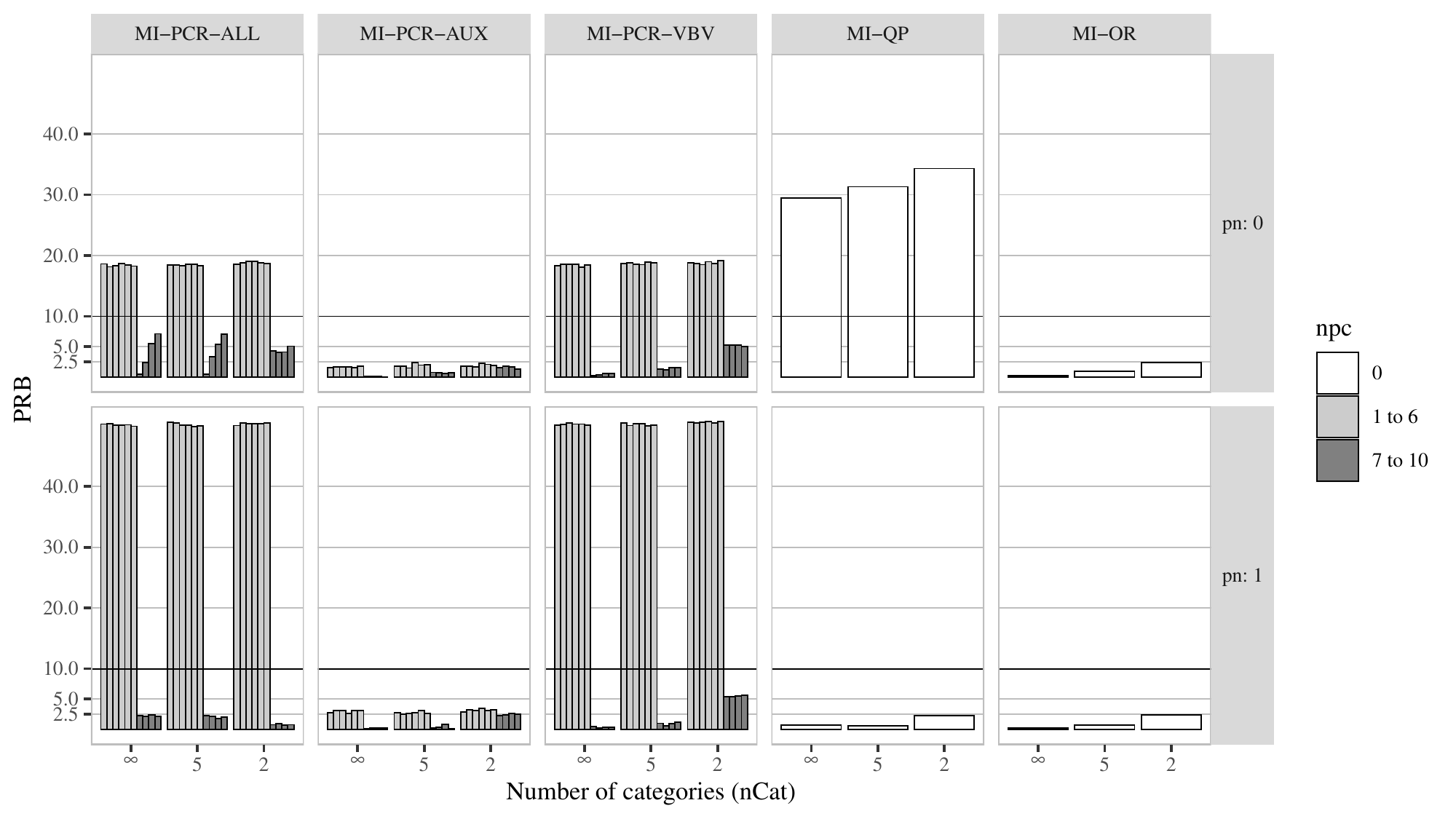} 

}

\end{knitrout}
\caption{\label{fig:biashd}
    Percent relative bias for the correlation between $\bm{x}_1$ and $\bm{x}_2$ in simulation study 2.
    $pn$ is the proportion of noise variables in $\bm{A}$.
    $npc$ is the number of PCs used by a given imputation method.
    The X-axis of each histogram distinguishes three levels of coarsening for the potential auxiliary variables ($nCat = (\infty, 5, 2)$).
    For each MI-PCR method, we reported a different vertical bar for each PRB obtained using a different number of PCs (from 1 to 10, from left to right).
	}
\end{figure}

\begin{figure}
\centering
\begin{knitrout}
\definecolor{shadecolor}{rgb}{0.969, 0.969, 0.969}\color{fgcolor}

{\centering \includegraphics[width=\maxwidth]{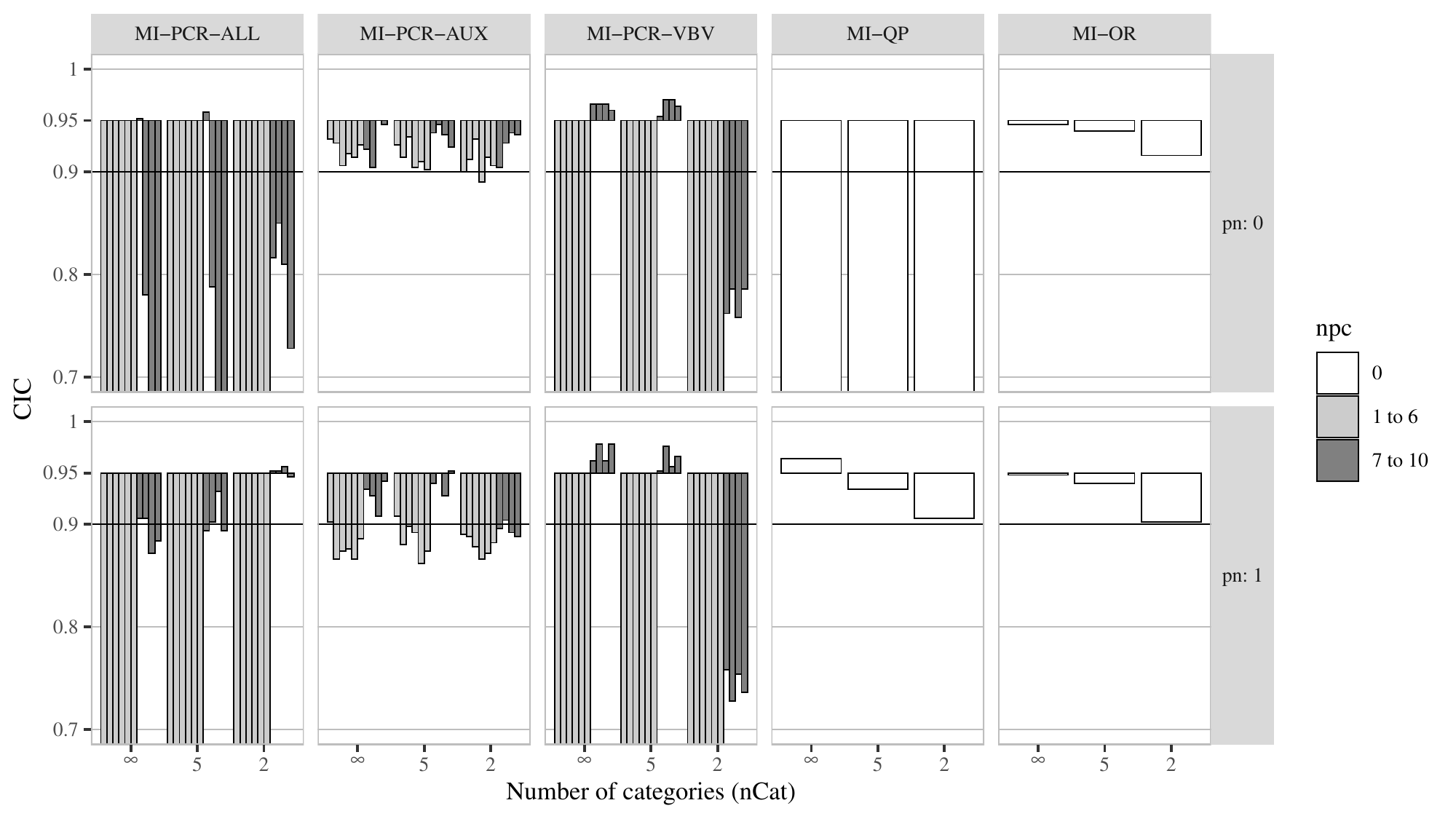} 

}

\end{knitrout}
\caption{\label{fig:cichd}
    Confidence interval coverage for the correlation between $\bm{x}_1$ and $\bm{x}_2$ in simulation study 2.
    $pn$ is the proportion of noise variables in $\bm{A}$.
    $npc$ is the number of PCs used by a given imputation method.
    The X-axis of each histogram distinguishes three levels of coarsening for the potential auxiliary variables ($nCat = (\infty, 5, 2)$).
    For each MI-PCR method, we reported a different vertical bar for each CIC obtained using a different number of PCs (from 1 to 10, from left to right).
	}
\end{figure}

\begin{figure}
\centering
\begin{knitrout}
\definecolor{shadecolor}{rgb}{0.969, 0.969, 0.969}\color{fgcolor}

{\centering \includegraphics[width=\maxwidth]{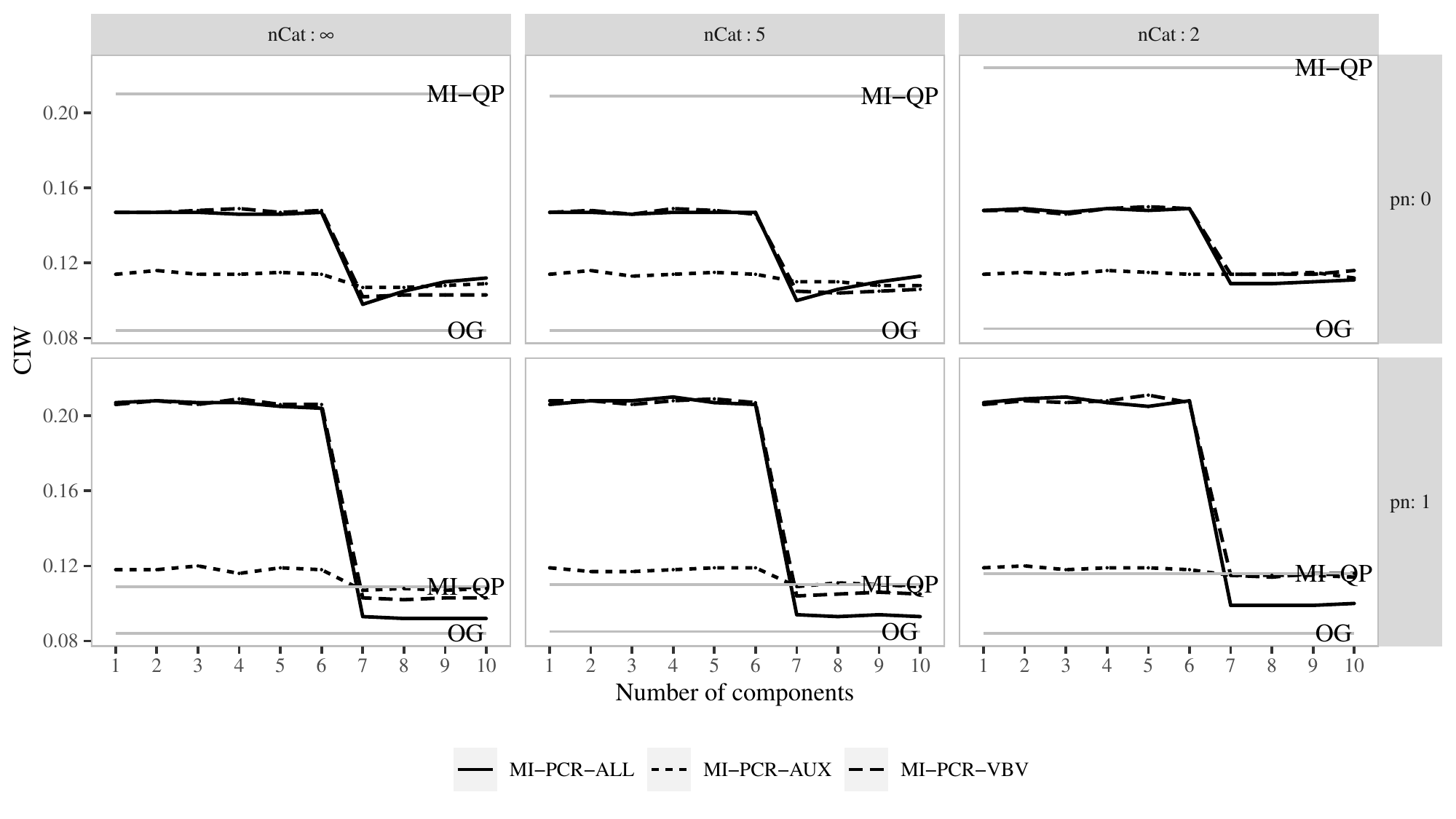} 

}

\end{knitrout}
\caption{\label{fig:ciwhd}
	Average confidence interval width for the correlation between $\bm{x}_1$ and $\bm{x}_2$ in simulation study 2.
    $nCat$ is the number of categories for the items in $\bm{M}$ and $\bm{A}$.
    $pn$ is the proportion of noise variables in $\bm{A}$.
	}
\end{figure}

In Figures~\ref{fig:biashd},~\ref{fig:cichd}, and~\ref{fig:ciwhd}, we reported the PRB, CIC, and CIW, respectively, for the correlation coefficient between $\bm{x}_1$ and $\bm{x}_2$ in an illustrative selection of conditions.
The same overall patterns described in the first simulation study were still present.
However, a few key differences did arise:
\begin{itemize}
\item In simulation study 1---with only 56 total predictors---MI-PCR-ALL and MI-PCR-VBV showed a gradual improvement in performance as more components were used.
  However, with $p = 242$, both methods showed a persistently high bias (PRB $> 10$) and low coverage (CIC $< 0.7$) when using fewer than 7 components.
  Both methods also demonstrated a more sudden improvement in performance at the $7$th PC mark.
\item The performance of MI-QP suffered relatively more from decreasing proportions of noise variables.
  In both studies, it was clear that MI-QP led to lower bias and better CIC when only a few predictors were correlated with the variables under imputation (i.e., for higher values of $pn$.)
  However, in simulation study 2, the increased collinearity due to $pn = 0$, resulted in extreme bias (PRB $> 20$) and under-coverage of the true parameter values (CIC $< 0.7$).
\end{itemize}

\begin{figure}
\centering
\begin{knitrout}
\definecolor{shadecolor}{rgb}{0.969, 0.969, 0.969}\color{fgcolor}

{\centering \includegraphics[width=\maxwidth]{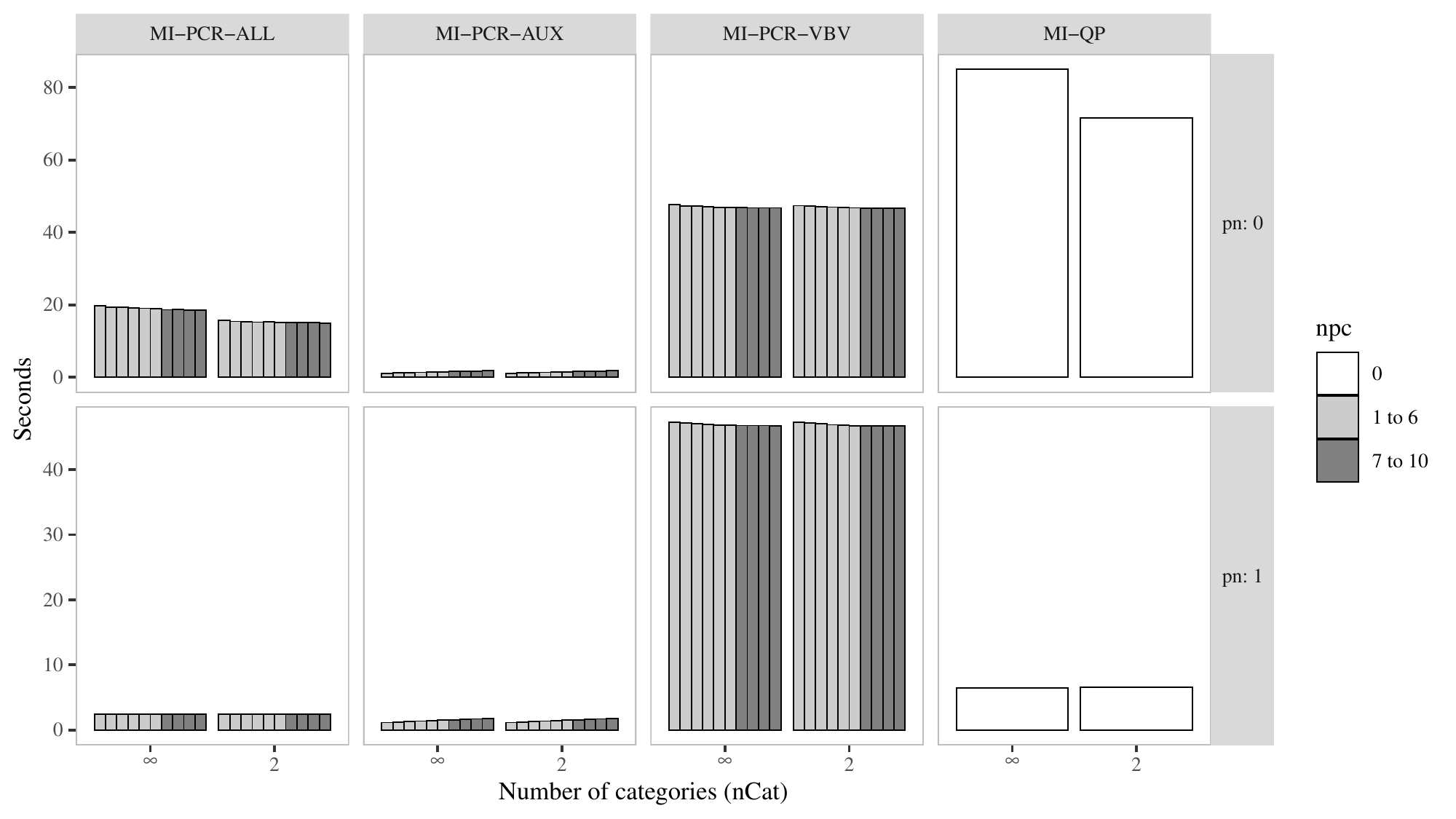} 

}

\end{knitrout}
\caption{\label{fig:timehd}
	Average imputation time in simulation study 2.
    $nCat$ is the number of categories for the items in $\bm{M}$ and $\bm{A}$.
    $pn$ is the proportion of noise variables in $\bm{A}$.
	}
\end{figure}

In Figure~\ref{fig:timehd} we reported the average imputation time in seconds for all imputation methods.
MI-PCR-AUX was the fastest, taking just a few seconds to run through the five chains and 20 iterations of the mice algorithm.
MI-PCR-VBV was the PCA-based method taking the longest time, with an average imputation time of around 40 seconds.
MI-PCR-ALL and MI-QP were impacted by the number of noise variables in the data.
Both took less than 10 seconds in the presence of many noise variables.
However, for $pn = 0$, they took around 20 and 80 seconds, respectively.
Both methods also took a few seconds less when the predictor data had been dichotomized.

% Read plots

% BIAS

% CONFIDENCE INTERVAL COVERAGE

% <<plot-cic-correlation-hd, cache = FALSE, echo = F>>=
% read_chunk("./code/plot-hd.R")
% @

% CONFIDENCE INTERVAL WIDTH

% <<plot-ciw-correlation-hd, cache = FALSE, echo = F>>=
% read_chunk("./code/plot-hd.R")
% @

% IMPUTATION TIME

% <<plot-time-hd, cache = FALSE, echo = F>>=
% read_chunk("./code/plot-hd.R")
% @

% Project:  paper-mipcr-compare
% Topic:    Present fireworks disaster data comparison
% Author:   Edoardo Costantini
% Created:  2021-12-20
% Modified: 2022-12-27

\section{Case study: fireworks disaster data}\label{sec:case-study}

To understand the performance of MI-PCR in real data, we compared the performance of the three MI-PCR implementations described above to an imputation carried out by an imputation expert on a real-world data set.
\citet[pp. 315--317]{vanBuuren:2018} gives a detailed description of how he solved the ``many variables'' problem for the Fireworks' disaster data set (FDD).
On May 13, 2000, the explosion of a fireworks storage facility in Enschede, the Netherlands, killed 23 people and injured approximately 950 others.
Many people residing in the neighborhood of the explosion experienced signs of post-traumatic stress disorder (PTSD).
The FDD was collected as part of a randomized controlled trial carried out in the aftermath of the explosion.
The data were collected to assess the efficacy of two treatments of anxiety-related disorders in children, in terms of reducing PTSD symptoms over time.
The main outcome measure for this analysis was the PTSD Reaction Index (PTSD-RI), measured as reported by the child and by the parent, at three different time points.
Fifty-two children were assigned either Eye Movement Desensitisation and Reprocessing (EMDR) treatment or Cognitive Behavioral Therapy (CBT).
Of the 65 variables recorded in the data set, 49 were incomplete.
The percentage of missing values on each variable ranged from 3\% to 50\%.
A complete-case analysis would have resulted in analyzing only eight cases.
To avoid the unacceptable reduction in sample size and biased parameter estimates, a principled missing data treatment was needed.

The major difficulty in imputing these data was the large number of predictors relative to the sample size ($p = 65, n = 52$).
To avoid over-parameterized imputation models, the imputation models' predictors needed to be carefully selected.
Van Buuren employed two main strategies to address the high-dimensional nature of the data:
\begin{itemize}
    \item Use only the first measurement of the outcomes as predictors in the imputation models of other outcomes.
        This choice reduced the number of predictors by two-thirds.
    \item Use the total scores of scales as predictors in the univariate imputation models of other variables, instead of the individual scale items.
            This was done using passive imputation (\citealp[p. 12]{vanBuurenOudshoorn:2000}; \citealp[pp. 1129--1130]{eekhoutEtAl:2018}).
\end{itemize}

Each of the three MI-PCR strategies considered in this report is suitable to address the large number of potential imputation model predictors in the FDD\@.
The advantage of MI-PCR is the automatic way in which large numbers of predictors can be accommodated.
Through this case study, we wish to illustrate the degree to which MI-PCR can produce results similar to those obtained by an expertly designed imputation procedure.
There are several reasons why the FDD is ideally suited to this purpose:
\begin{itemize}
    \item The data have key characteristics of social and behavioral science data sets (e.g., composite scales, longitudinal data).
    \item The code Van Buuren used to perform the MI procedure is freely accessible.
    \footnote{\url{https://github.com/stefvanbuuren/fimdbook/blob/master/R/fimd.R}}
    \item The data are freely accessible through the \emph{mice} R package~\citep{mice}.
    \item The reasoning behind the expert's MI procedure is well documented~\citep[p. 313]{vanBuuren:2018}.
\end{itemize}

\subsection{Method}\label{subsubsec:case-method}
The main analysis reported in \citet[p. 313]{vanBuuren:2018} focused on the effect of treatment on the mean PTSD-RI scores (both child-reported and parent-reported) across three time points.
Therefore, six variables were analyzed: the PTSD-RI total scores reported by children and their parents at three different time points.
We imputed these six analysis variables with the three MI-PCR procedures evaluated above and compared the pooled means obtained thereby with the results of Van Buuren's imputation procedure.
To evaluate the variability of the imputation methods, we repeated this procedure with 20 different random number seeds.

The results of the simulation study suggested that MI-PCR performs poorly if an insufficient number of components is extracted, whereas its performance is not severely impacted by selecting too many components.
Therefore, to choose the number of PCs in this case study, we decided to use the maximum number of PCs allowed by each imputation procedure.
The two variables with the most missing values had 25 cases observed.
Hence, at most 24 predictors could be used for each imputation model, and we could use at most 24 PCs.

We specified MI-PCR-AUX to impute the six items under analysis.
In each of the six univariate imputation models, MI-PCR-AUX used as predictors the other five variables under imputation and the first 19 PCs estimated from the remaining auxiliary columns (5 + 19 = 24).
We performed single imputation on the potential auxiliary variables to allow the PC estimation.
We used predictive mean matching as univariate imputation method and kept the imputations obtained after 20 iterations.
The predictors for the imputation models were selected by correlation-thresholding, with the threshold set to $0.1$.

The MI-PCR-ALL method used the first 24 PCs estimated on the entire data set---including the 6 analysis variables---as the sole predictors in each of the six univariate imputation models.
The same single imputation specification used for MI-PCR-AUX was used here to generate the complete data from which to estimate the PCs\@.
Finally, MI-PCR-VBV was performed by extracting 24 components from all the variables not under imputation for each univariate imputation model, at every iteration of the mice algorithm.

All starting imputations were created by a simple random draw from the data.
Convergence of the pre-processing single imputation used for MI-PCR-AUX and MI-PCR-ALL was assessed with the trace plots of the imputed values' means and standard deviations produced by the \texttt{plot.mids()} function from the \emph{mice} package.
We assessed the convergence of all the imputation methods with the same technique. 
We considered all methods to have converged within 20 iterations.
The plots of convergence trends can be viewed in the supplementary material.

\subsection{Results}\label{subsubsec:case-results}
\begin{figure}
\centering
\begin{knitrout}
\definecolor{shadecolor}{rgb}{0.969, 0.969, 0.969}\color{fgcolor}

{\centering \includegraphics[width=\maxwidth]{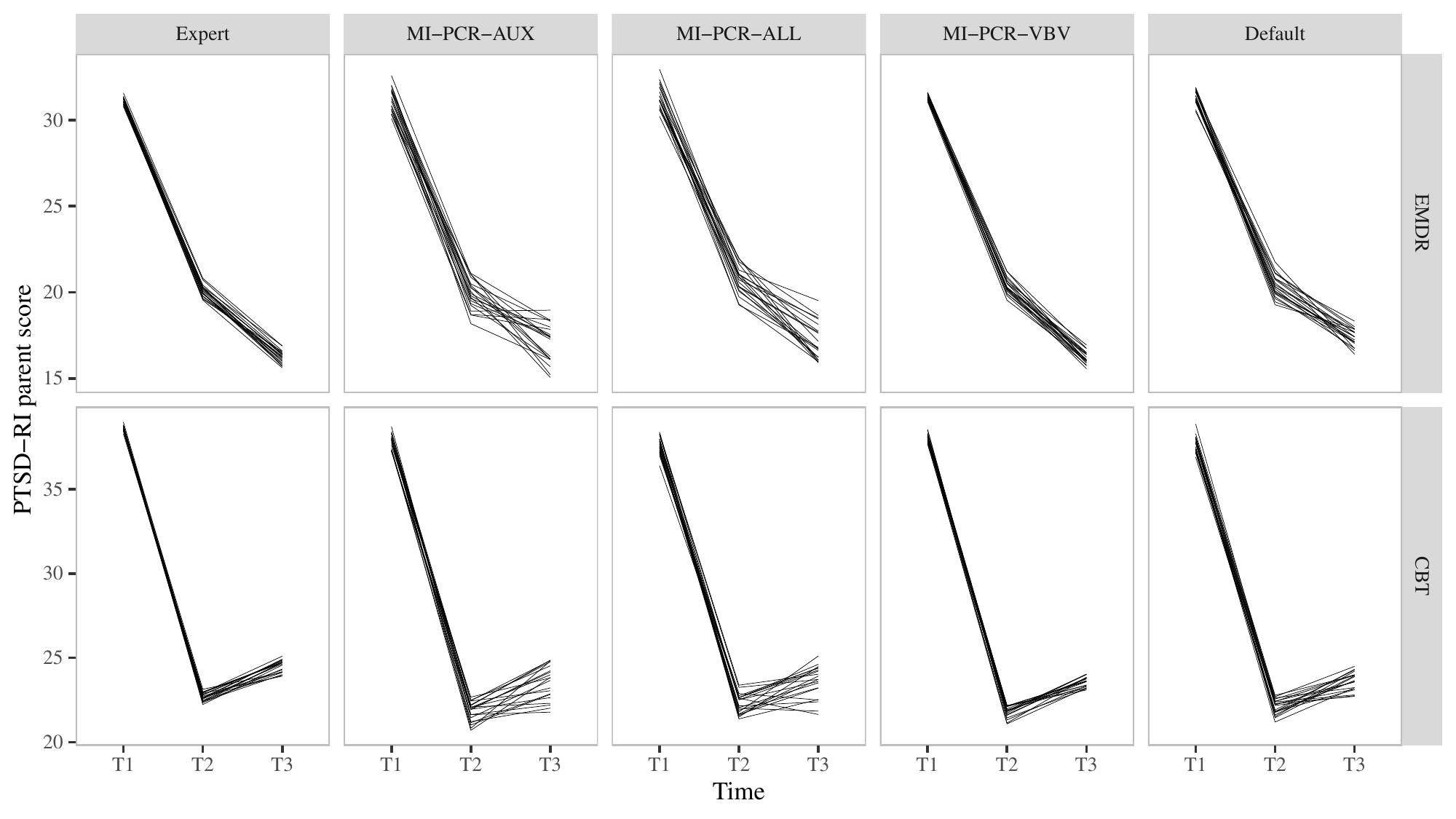} 

}

\end{knitrout}
\caption{\label{fig:fdd-yp}
	Mean levels of PTSD-RI parent score after imputation.
    The multiple lines plotted for each method represent results obtained with 20
    different seeds.
	}
\end{figure}

\begin{figure}
\centering
\begin{knitrout}
\definecolor{shadecolor}{rgb}{0.969, 0.969, 0.969}\color{fgcolor}

{\centering \includegraphics[width=\maxwidth]{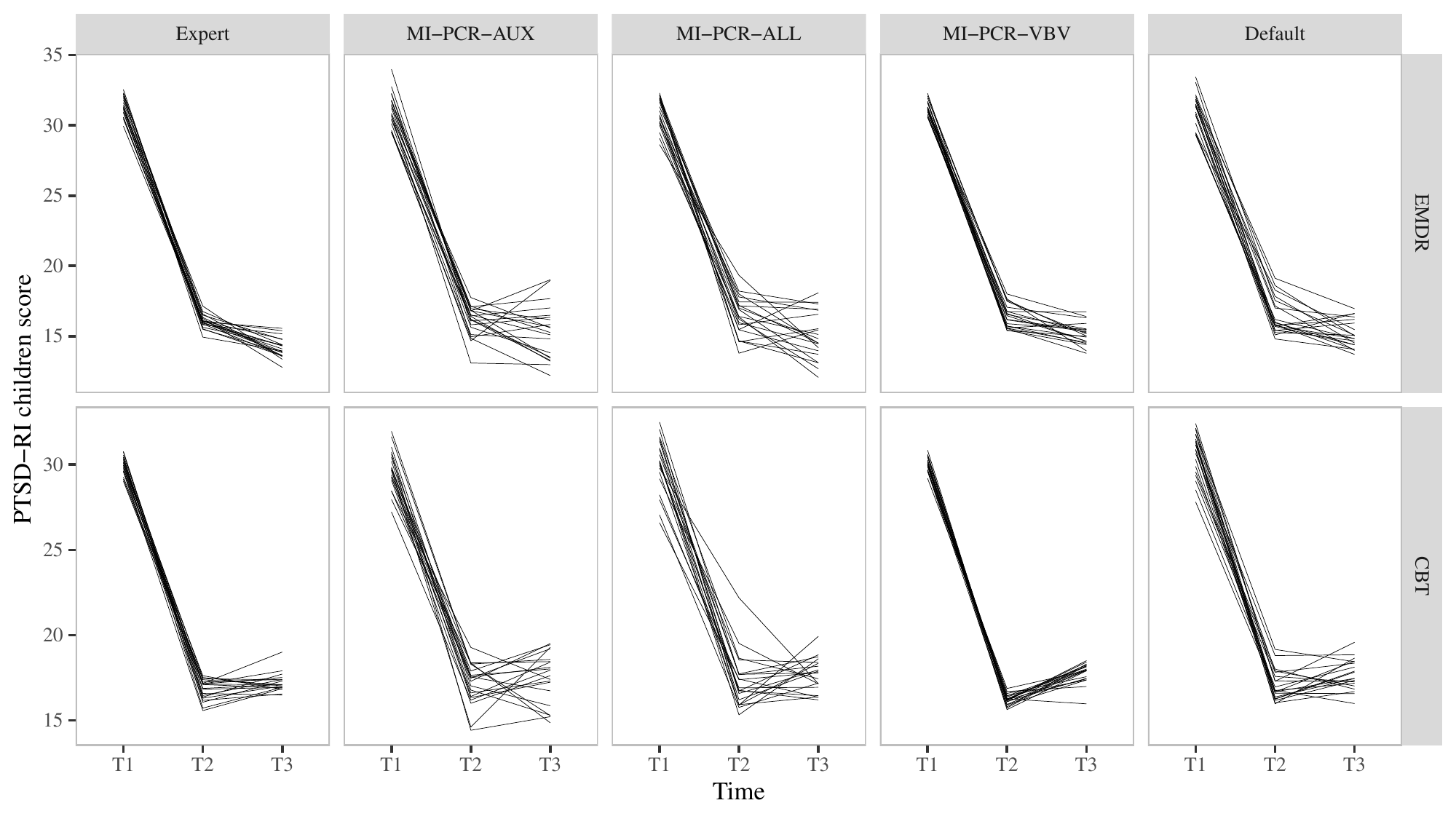} 

}

\end{knitrout}
\caption{\label{fig:fdd-yc}
	Mean levels of PTSD-RI children score after imputation.
    The multiple lines plotted for each method represent results obtained with 20
    different seeds.
	}
\end{figure}

Figures~\ref{fig:fdd-yp} and~\ref{fig:fdd-yc} report the (pooled) mean level of PTSD-RI over the three time points after imputation with the following approaches:
\begin{itemize}
    \item Expert's imputation models;
    \item The three MI-PCR approaches;
    \item Default run of \emph{mice} without any pre-processing and using all default argument values.
\end{itemize}

While all methods led to similar trends, variability of the imputations was noticeably higher for MI-PCR-AUX, MI-PCR-ALL, and the default run of \emph{mice}.
At every time point, the 20 different pooled means of the analyzed variables spread over a wider range of values compared to the expert's imputation.
This pattern held for both outcome variables, but it was most conspicuous for the child-reported PTSD-RI\@.
MI-PCR-AUX and MI-PCR-ALL had lower imputation precision than even the default run of \emph{mice} in this setup.
The performance of MI-PCR-VBV was on par with that of the expert's imputation: both methods produced comparable location and spread of the outcome variables pooled means.

% Load the plots for Fireworks disaster study

% Parent results

% Child results

% Project:  paper-mipcr-compare
% Topic:    General Discussion
% Author:   Edoardo Costantini
% Created:  2021-12-17
% Modified: 2023-01-09

\section{General Discussion}\label{sec:discussion}

In this study, we were interested in understanding the performance of the various MI-PCR methods as a function of the number of components used, the coarseness of the predictor data, and the amount of noise in the data.
Our findings suggest that MI-PCR performs well across a wide range of conditions, and MI-PCR-VBV shows the best performance of the three implementations we evaluated.
So long as the number of PCs met or exceeded the true number of latent variables, MI-PCR-VBV outperformed the standard correlation-thresholding approach (MI-QP) on all the metrics we considered.
Furthermore, the good imputation performance of MI-PCR-VBV comes with some desirable features that are missing from the other MI-PCR methods.
First, MI-PCR-VBV does not rely on knowledge of the analysis model, although such knowledge is easily incorporated when available.
Second, MI-PCR-VBV does not require a pre-processing single imputation step.
Third, when MI-PCR-VBV resulted in deviations from nominal coverage, it tended towards over-coverage.
Finally, in the case study, MI-PCR-VBV was able to automatically obtain results that were essentially indistinguishable from those produced by an expertly designed imputation model.

The good performance of MI-PCR-VBV comes at the expense of computation time.
Performing PCA for every variable under imputation at every iteration of the MICE algorithm requires a much larger number of computations than the other two MI-PCR methods which leads to a drastically higher imputation time.
As the number of variables to impute increases, computation time might become excessive.
This makes the MI-PCR-VBV strategy more suitable for broad and intermediate imputation scopes \citep[][p. 46]{vanBuuren:2018}, where imputation is performed once by an institution with adequate computational resources, and then delivered to a collection of researchers to be used in their analysis.
    
% Project:  paper-mipcr-compare
% Topic:    Supplementary material: simulation study 1
% Author:   Edoardo Costantini
% Created:  2023-03-06
% Modified: 2023-03-18

\section{Some final remarks on the number of components}\label{sec:final-remarks-npcs}

% Read all plots for the low dimensional condition (supplementary material)

% Describe procedure

In practice, imputers using MI-PCR need a decision rule to select the number of components.
Using the same number of components as the total available variables, as we did in Section~\ref{sec:case-study}, is not a viable solution for data sets with hundreds of variables (or more).
Fortunately, a variety of decision rules have been proposed for this purpose \citep{zwickVelicer:1986}.
Based on the results described in Sections~\ref{sec:simulation-study} and~\ref{sec:simulation-study-2}, we can infer that any decision rule that selects the ``true'' number of components, or more, would produce satisfactory results with MI-PCA.

To gain some preliminary insight into how these decision rules could impact the performance of MI-PCA, we applied four non-graphical decision rules\footnote{We used the implementation of these rules provided by the `nScree()' function in the R package \textit{nFactor}~\citep{nFactors}.} described by~\cite{raicheEtAl:2013} to 500 data sets generated according to the procedure described in Sections~\ref{sec:simulation-study} and~\ref{sec:simulation-study-2}.
Specifically, we implemented the optimal coordinates index (\textit{oc}), the acceleration factor (\textit{af}), the Kaiser criterion (\textit{kc}), and the parallel analysis criterion (\textit{pa}) using both the fully observed, discretized data and the complete cases available after imposing the missing values.

In Table~\ref{tab:ngdr}, we report the lowest, the highest, and the median number of principal components retained with each decision rule across the 500 data sets.
For each decision rule, the number of PCs selected when analyzing the complete cases was always equal to or higher than the number of PCs selected when analyzing the fully observed data.
The \textit{oc} criterion demonstrated mixed performance.
The median number of PCs selected by \textit{oc} was always at least 7, but the minimum number of components selected was less than 7 in all conditions.
The \textit{kc} and \textit{pa} decision rules selected between 7 and 15 PCs when applied to data sets with $P = 56$ columns and between 7 and 39 PCs when applied to data with $P = 242$ columns.
Hence, \textit{kc} and \textit{pa} appear to be safe options when considering our desideratum of selecting no fewer than the true number of components. 
In line with results presented by~\cite{raicheEtAl:2013}, \textit{af} underestimated the number of PCs to retain and always selected fewer than 7 PCs.
Based on these results, we can tentatively suggest applying the Kaiser criterion or the parallel analysis criterion to the complete case as a viable method of selecting the number of PCs to use in MI-PCA.

% latex table generated in R 4.2.2 by xtable 1.8-4 package
% Thu Mar 23 09:35:03 2023
\begin{table}[ht]
\centering
\scalebox{.65}{
\begin{tabular}{|rr|cccc|cccc|cccc|cccc|cccc|cccc|}
  \hline \rowcolor{gray!20} & & \multicolumn{24}{c|}{P = 56} \\
  \hline \rowcolor{gray!20} & data & \multicolumn{12}{c|}{Fully observed} & \multicolumn{12}{c|}{Complete cases}\\
 \hline \rowcolor{gray!20} & pn  & \multicolumn{4}{c|}{0} & \multicolumn{4}{c|}{0.67} & \multicolumn{4}{c|}{1} & \multicolumn{4}{c|}{0} & \multicolumn{4}{c|}{0.67} & \multicolumn{4}{c|}{1}\\
 \hline \rowcolor{gray!20} & rule & oc & af & kc & pa & oc & af & kc & pa & oc & af & kc & pa & oc & af & kc & pa & oc & af & kc & pa & oc & af & kc & pa\\
 \hline
\multirow{3}{*}{nCat = $\infty$} & highest & 7 & \textbf{1} & 7 & 7 & 7 & \textbf{1} & 7 & 7 & 7 & 7 & 7 & 7 & 7 & \textbf{1} & 7 & 7 & 7 & 7 & 7 & 7 & 7 & 7 & 7 & 7 \\ 
   & median & 7 & \textbf{1} & 7 & 7 & 7 & \textbf{1} & 7 & 7 & 7 & 7 & 7 & 7 & 7 & \textbf{1} & 7 & 7 & 7 & \textbf{1} & 7 & 7 & 7 & 7 & 7 & 7 \\ 
   & lowest & \textbf{1} & \textbf{1} & 7 & 7 & \textbf{1} & \textbf{1} & 7 & 7 & \textbf{1} & \textbf{1} & 7 & 7 & \textbf{1} & \textbf{1} & 7 & 7 & \textbf{2} & \textbf{1} & 7 & 7 & \textbf{1} & \textbf{1} & 7 & 7 \\ 
   \hline
\multirow{3}{*}{nCat = 5} & highest & 7 & \textbf{1} & 7 & 7 & 7 & \textbf{1} & 7 & 7 & 7 & 7 & 7 & 7 & 8 & \textbf{1} & 8 & 8 & 8 & 7 & 8 & 8 & 7 & 7 & 7 & 7 \\ 
   & median & 7 & \textbf{1} & 7 & 7 & 7 & \textbf{1} & 7 & 7 & 7 & 7 & 7 & 7 & 7 & \textbf{1} & 7 & 7 & 7 & \textbf{1} & 7 & 7 & 7 & 7 & 7 & 7 \\ 
   & lowest & \textbf{1} & \textbf{1} & 7 & 7 & \textbf{2} & \textbf{1} & 7 & 7 & \textbf{1} & \textbf{1} & 7 & 7 & \textbf{1} & \textbf{1} & 7 & 7 & \textbf{1} & \textbf{1} & 7 & 7 & \textbf{1} & \textbf{1} & 7 & 7 \\ 
   \hline
\multirow{3}{*}{nCat = 2} & highest & 8 & \textbf{1} & 8 & 8 & 8 & \textbf{1} & 8 & 8 & 8 & 7 & 8 & 8 & 14 & \textbf{1} & 15 & 15 & 13 & 7 & 13 & 13 & 13 & 7 & 13 & 13 \\ 
   & median & 7 & \textbf{1} & 7 & 7 & 7 & \textbf{1} & 7 & 7 & 7 & 7 & 7 & 7 & 11 & \textbf{1} & 12 & 12 & 10 & \textbf{1} & 11 & 11 & 10 & 7 & 10 & 10 \\ 
   & lowest & \textbf{1} & \textbf{1} & 7 & 7 & \textbf{1} & \textbf{1} & 7 & 7 & \textbf{1} & \textbf{1} & 7 & 7 & \textbf{1} & \textbf{1} & 9 & 9 & \textbf{1} & \textbf{1} & 8 & 8 & \textbf{1} & \textbf{1} & 8 & 8 \\ 
   \hline \rowcolor{gray!20} & & \multicolumn{24}{c|}{P = 242} \\
 \hline
\multirow{3}{*}{nCat = $\infty$} & highest & 7 & \textbf{1} & 7 & 7 & 7 & \textbf{5} & 7 & 7 & 7 & \textbf{6} & 7 & 7 & 35 & \textbf{1} & 35 & 35 & 28 & \textbf{6} & 29 & 29 & 25 & \textbf{6} & 25 & 25 \\ 
   & median & 7 & \textbf{1} & 7 & 7 & 7 & \textbf{1} & 7 & 7 & 7 & \textbf{1} & 7 & 7 & 21 & \textbf{1} & 25 & 25 & 19 & \textbf{1} & 21 & 21 & 18 & \textbf{6} & 19 & 19 \\ 
   & lowest & \textbf{1} & \textbf{1} & 7 & 7 & \textbf{2} & \textbf{1} & 7 & 7 & \textbf{1} & \textbf{1} & 7 & 7 & \textbf{2} & \textbf{1} & 14 & 14 & \textbf{2} & \textbf{1} & 14 & 14 & \textbf{1} & \textbf{1} & 14 & 14 \\ 
   \hline
\multirow{3}{*}{nCat = 5} & highest & 7 & \textbf{1} & 7 & 7 & 7 & \textbf{5} & 7 & 7 & 7 & \textbf{6} & 7 & 7 & 39 & \textbf{1} & 39 & 39 & 32 & \textbf{6} & 32 & 32 & 27 & \textbf{6} & 31 & 31 \\ 
   & median & 7 & \textbf{1} & 7 & 7 & 7 & \textbf{1} & 7 & 7 & 7 & \textbf{1} & 7 & 7 & 23 & \textbf{1} & 28 & 28 & 22 & \textbf{1} & 24 & 24 & 20 & \textbf{6} & 22 & 22 \\ 
   & lowest & \textbf{1} & \textbf{1} & 7 & 7 & \textbf{2} & \textbf{1} & 7 & 7 & \textbf{1} & \textbf{1} & 7 & 7 & \textbf{3} & \textbf{1} & 19 & 19 & \textbf{3} & \textbf{1} & 17 & 17 & \textbf{3} & \textbf{1} & 16 & 16 \\ 
   \hline
\multirow{3}{*}{nCat = 2} & highest & 13 & \textbf{1} & 13 & 13 & 13 & \textbf{5} & 13 & 13 & 13 & \textbf{6} & 13 & 13 & 41 & \textbf{1} & 41 & 41 & 35 & \textbf{6} & 39 & 39 & 34 & \textbf{6} & 34 & 34 \\ 
   & median & 10 & \textbf{1} & 10 & 10 & 10 & \textbf{5} & 10 & 10 & 10 & \textbf{1} & 10 & 10 & 29 & \textbf{1} & 34 & 34 & 28 & \textbf{5} & 30 & 30 & 26 & \textbf{6} & 29 & 29 \\ 
   & lowest & \textbf{1} & \textbf{1} & 8 & 8 & \textbf{2} & \textbf{1} & 8 & 8 & \textbf{1} & \textbf{1} & 8 & 8 & \textbf{1} & \textbf{1} & 26 & 26 & \textbf{1} & \textbf{1} & 24 & 24 & \textbf{3} & \textbf{1} & 22 & 22 \\ 
   \hline
\end{tabular}
}
\caption{The lowest, the highest, and the median number of principal components selected by four non-graphical decision rules across 500 data sets generated according to the simulation design described in Sections \ref{sec:simulation-study} and \ref{sec:simulation-study-2}. The decision rules reported are the optimal coordinates index ($oc$), the acceleration factor ($af$) the Kaiser criterion ($kc$), and the parallel analysis criterion ($pa$). The table distinguishes between the results obtained when applying the four decision rules to two types of data (the originally fully observed data and the complete cases), data with two different dimensionalities ($P = 56, 242$), data with different discretization levels ($nCat = \infty, 5, 2$), and data with different proportions of noise variables ($pn = 0, 0.67, 1$). Numbers below 7 are reported in bold.} 
\label{tab:ngdr}
\end{table}

% Project:  paper-mipcr-compare
% Topic:    Limitations and future directions
% Author:   Edoardo Costantini
% Created:  2021-12-17
% Modified: 2023-03-18

\section{Limitations and future directions}\label{sec:limitations}

In the preceding section, we provide some preliminary insight into how four non-graphical PC enumeration methods might affect the performance of MI-PCR, but our results support only tentative recommendations.
Future research should focus specifically on the issue of PC enumeration in MI-PCA and thoroughly explore which rule is most suitable for the imputation task.
Furthermore, the unsupervised nature of PCA introduces an additional dimension into the decision calculus.
The MI-PCR implementations we compared here extract PCs without considering the relationship between the imputation model's predictors and outcome.
Consequently, the MI-PCR methods we evaluated could, potentially, extract components that explain relatively little variance in the variables under imputation, regardless of how many PCs are retained.
To mitigate this possibility, MI-PCR-VBV could be implemented with some form of supervision, such as supervised PCA \citep{bairEtAl:2006} or principal covariates regression \citep{deJongKiers:1992}.
In addition to avoiding poorly predictive PCs, a supervised version of MI-PCR might require fewer components to obtain the same imputation quality.
We are currently exploring these possibilities in a follow-up to the study reported here.

It is common for social scientists to analyze non-normal data such as ordinal rating scales (e.g., any item in the NEO-FFI and NEO-PI-R) or skewed social variables (e.g., items affected by extreme response styles).
The results of this study apply directly to situations wherein the non-normal variables are the targets of imputation.
MI-PCR only adjusts the right-hand-side of the univariate imputation models.
So, our approach can be directly applied to univariate imputation models for non-normal data (e.g., Bayesian logistic and polytomous regression models, predictive mean matching).
However, using PCA to extract components from a set of non-normal predictors requires more careful consideration.
In general, as a tool to summarize variation on a set of variables, PCA does not need to meet rigorous distributional assumptions \citep[][pp. 19, 49, 338]{jolliffe:2002}.
However, when the variables in $\bm{X}$ deviate from normality because of the presence of extreme cases, robust PCA can be used to reduce the impact these observations have on the estimation of the PCs.
For example, \cite{crouxRuiz-Gazen:2005} and \cite{hubertEtAl:2009} proposed alternative PCA computations that are robust to outliers and asymmetry.
Similarly, PCAMIX~\citep{kiers:1991, chaventEtAl:2012} can be used in the presence of a mix of continuous, ordinal, and nominal variables.
Luckily, MI-PCR has a modular structure, and the classical PCA estimation we applied in this study can be replaced by any alternative PCA approach.
These alternatives could improve the performance of MI-PCR when it is applied to non-normal data, but we do not expect this change to have an impact on the relative performances of the different MI-PCR implementations we evaluated here.
Any possible improvement in the quality of the PCs would impact each implementation equally, so our overall conclusions would be unlikely to change.
Nevertheless, it would be interesting to evaluate the extent of the improvement that could be achieved by incorporating more robust versions of PCA.

The missing data mechanism studied in this paper is relatively simple.
The probability of missing values on $\bm{T}$ depends, through a logit function, on the linear combination of the predictors $\bm{M}$.
However, interactions and polynomial terms might be present in the linear component of Equation \ref{eq:logit}.
MI-PCR can address this complexity by augmenting the set of variables from which PCs are extracted with all the interaction and polynomial terms of interest.
To what extent this strategy is feasible and effective remains to be explored.
Furthermore, while we focused on right tail MAR, \cite{schoutenVink:2021} and \cite{collinsEtAl:2001} have shown that the `shape' of the missing data mechanism has an impact on the severity of a missing data problem.
In the worst case, we expect the different MAR shapes to impact the absolute performance of the MI-PCR methods compared in this study, but not their relative performance.
Future research could address this issue in detail.

Multilevel data provide a similar avenue for future research.
Social science data are often characterized by clusters of observations.
Imputation procedures that ignore this feature of the data can lead to biased estimates as imputations are generated without considering cluster dependencies~\citep{reiterEtAl:2006}.
One way to address this issue is to include dummy variables representing the cluster effects in the imputation models~\citep[fixed effect imputation,][]{enders:2016}.
However, this approach has the disadvantage of increasing the dimensionality of the design matrix to an impractical extent.
Using PCs to reduce dimensionality might be a good way to address grouping in the data without incurring estimation difficulties of sophisticated multilevel imputation procedures.

Finally, the good performance of MI-PCR-ALL in the conditions with dichotomized auxiliary variables remains something of a mystery.
We did not expect this pattern, and we do not have an explanation for this finding.
Since the other MI-PCR methods performed at their worst in the $nCat = 2$ condition, it would be interesting to further explore the capabilities of MI-PCR-ALL in this special case.
    
% Project:  paper-mipcr-compare
% Topic:    Conclusions
% Author:   Edoardo Costantini & Kyle M. Lang
% Created:  2022-12-27
% Modified: 2023-01-09

\section{Conclusions}\label{sec:conclusions}

This study extends and refines the findings of \citet{costantiniEtAl:2022a} by providing further information on how to best incorporate PCA into MI.
In our simulation studies, using PCR as a univariate imputation method within every iteration of a MICE algorithm (i.e., MI-PCR-VBV) provided small bias, good statistical efficiency, and close to nominal coverage.
Our case study added to these findings by showing that MI-PCR-VBV can provide performance on-par with expertly designed imputation.
Although computational demand could become a limiting factor in some situations, our findings suggest that MI-PCR-VBV is a promising general-purpose, imputation algorithm that can streamline the process of conducting principled MI in data sets with many variables.
    
    % Stop section header numbering here
    \setcounter{secnumdepth}{4}

    % Extra sections required by the journal
    \section{Funding}

    No funds, grants, or other support were received.

    \section{Conflicts}

    The authors have no relevant financial or non-financial interests to disclose.

    \section{Availability of data and materials}

    The data used for the case study is publicly available as part of the R package \textit{mice} under the name \textit{fdd} [https://github.com/amices/mice/blob/master/data/fdd.rda].

    \section{Code availability}

    The code used for the simulation study is available on Zenodo \citep{costantini:2023a}.
    The code used for the case study is available on Zenodo \citep{costantini:2023b}.
    Please read the README.md files for instructions on how to replicate the results.

    % References
    \bibliography{\pathBIB/bibshelf}

\end{document}